\documentclass[11pt]{article}

\usepackage{geometry,showlabels, fullpage, authblk}
\usepackage{xr-hyper}
\usepackage{times,enumitem}
\usepackage{hyperref}[]
\hypersetup{
	colorlinks=true,
	linkcolor=blue,
	filecolor=magenta,      
	urlcolor=cyan,
	citecolor=blue,
}
\usepackage{ulem}

\usepackage{url}

\usepackage{bigints}
\usepackage{amsfonts,amsmath,amssymb,amsthm, bbm}
\usepackage{wrapfig}
\usepackage{verbatim,float,url}
\usepackage{graphicx}
\usepackage{subcaption}
\usepackage[round]{natbib}
\usepackage[english]{babel}
\usepackage{cancel}
\usepackage{color}
\usepackage[dvipsnames]{xcolor}
\usepackage{cleveref}
\usepackage{xspace}
\usepackage{multirow}

\usepackage{collcell}
\makeatletter
\newcolumntype{G}{>{\collectcell\@gobble}c<{\endcollectcell}@{}}
\makeatother

\usepackage{bm}

\usepackage{mathtools}

\newtheorem{theorem}{Theorem}
\newtheorem{lemma}[theorem]{Lemma}

\usepackage{mdwlist}

\theoremstyle{definition}

\usepackage{tikz}  
\usetikzlibrary{trees}
\usetikzlibrary{arrows.meta}
\usetikzlibrary{decorations.pathreplacing}

\usepackage{mathtools}

\DeclarePairedDelimiter\floor{\lfloor}{\rfloor}

\def\E{\mathbb{E}}

\graphicspath{{figs/}}

\newcommand{\iidsim}{\overset{iid}{\sim}} % iid simulated
\newcommand{\iid}{\rm independent and identically distributed }

\usepackage{ulem}
\usepackage{xspace} %SR added for \sr

\usepackage{algorithm}% http://ctan.org/pkg/algorithms
\usepackage{algpseudocode}% http://ctan.org/pkg/algorithmicx

\algnewcommand\algorithmicinput{\textbf{INPUT:}}
\algnewcommand\INPUT{\item[\algorithmicinput]}
\algnewcommand\algorithmicoutput{\textbf{OUTPUT:}}
\algnewcommand\OUTPUT{\item[\algorithmicoutput]}

\DeclareMathOperator*{\argmax}{arg\,max}

\theoremstyle{plain}
\newtheorem{assumption}{Assumption}
\allowdisplaybreaks
\usepackage{bm}

\title{Scalable Bayesian change point detection with spike and slab priors}

\author[1]{Lorenzo Cappello}
\author[2]{Oscar Hernan Madrid Padilla}
\author[3]{Julia A. Palacios}

\affil[1,3]{\small Department of Statistics, Stanford University}
\affil[2]{\small Department of Statistics, University of California, Los Angeles}
\affil[3]{\small Department of Biomedical Data Science, Stanford University}

\begin{document}
	\maketitle
	
	\begin{abstract}
We study the use of spike and slab priors for consistent estimation of the number of change points and their locations. Leveraging recent results in the variable selection literature, we show that an estimator based on spike and slab priors achieves optimal localization rate in the multiple offline change point detection problem. Based on this estimator, we propose a Bayesian change point detection method, which is one of the fastest Bayesian methodologies, and it is more robust to misspecification of the error terms than the competing methods. We demonstrate through empirical work the good performance of our approach vis-a-vis some state-of-the-art benchmarks. 
	
		\vskip 5mm
		\textbf{Keywords}:  Optimality, shrinkage, robust, approximate inference
	\end{abstract}
	
	%\section{Introduction}
	
\section{Introduction}
\label{sec:intro}

Change point detection has received considerable attention in the statistical literature for several decades. Assume we observe a vector of  independent random variables $\textbf{Y}=(Y_1,\ldots,Y_T)^{\top}$ according to the linear model
\begin{equation}
	\label{eq:model}
	Y_t = f_t + \epsilon_t \,\,\,\,\,\,\, \text{for } 1\leq  t \leq T,
\end{equation}
where $f_t$ is a right continuous function with an unknown number $K$ of change points, and $\epsilon_t$'s  are \iid random variables with $\E [\epsilon_t]=0$ for all $t$. The main goal of  offline change-point detection is to simultaneously estimate $K$ and the locations of the change points of $f_t$. A natural assumption is that $f_t$ (i.e., the conditional mean) is sparse, in the sense that there are only a small number of change-points. We further assume that $f_t$ is piecewise-constant (i.e. a right continuous step function).  %We can write this as 
%\begin{equation}
%	\label{eqn:model}
%	f_t \,=\, \begin{cases}
%		\mu_0   &   \text{if }\,\,\,\,  t  \leq \eta_1-1\\
%		\mu_1   &   \text{if }\,\,\,\,  \eta_1  \leq t  \leq \eta_2-1\\
%		\vdots\\
%		\mu_K & \text{if }   \,\,\,\,  \eta_K \leq t  \leq T,\\
%	\end{cases}
%\end{equation}
These modeling choices lead to a low-dimensional parametric model that is interpretable, can fit non-stationary time-series, and is suitable for prediction. 

The literature on change point detection includes a large number of frequentist methods. Most of these methodologies rely on a test statistic to detect parametric changes in the distribution of the observables and model selection techniques to determine the number of parameters defining the signal $f_t$. Some examples of test statistics include the likelihood ratio and the CUSUM statistic \citep{pag54}.
  The model selection step typically relies on either an $\ell_0$ or an $\ell_1$ penalty. The penalty is included directly through a penalized likelihood or via an information criterion such as AIC and BIC. Consequently, several variable selection methodologies have a conceptual analogue in change point detection: for example, the Dantzig selector of \cite{can07} has the same rationale as the multiscale SMUCE estimator of \cite{fric14}; the total variation denoising \citep{rud92} and the fused LASSO procedure \citep{tib05} share the same penalty with the LASSO \citep{tib96}. Other frequentist methods for univariate change point detection include the wild  binary segmentation of \cite{fryzlewicz2014wild} based on the CUSUM statistics, and various algorithms for $\ell_0$ penalized change point detection \citep{friedrich2008complexity,rigaill2010pruned,killick2012optimal,maidstone2017optimal}.

From the Bayesian perspective, popular change point detection methods rely on product partition models \citep{bar92,bar93}. However, the use of MCMC to approximate the posterior distributions of these models is challenging \citep{chi96,chi98}, and much research has focused on alternatives to MCMC: \cite{fea06} proposed two algorithms to perform direct simulation from the posterior distribution (one to do exact simulation from the posterior and one using an approximate version); \cite{rig12} derived exact formulae for the posterior distribution. Recent works take an empirical Bayes approach to set the prior distributions \citep{du16, liu20}. \cite{liu20} is more general, allowing to recover piecewise polynomial signals. To the best of our knowledge, Bayesian variable selection procedures %have not been as widely employed for multiple change points detection as it has happened in the frequentist literature. Sparsity inducing priors
 such as the horseshoe prior \citep{car10} and the spike and slab prior \citep{mit88}, have not been studied in this setting. Recent works used the horseshoe prior for trend filtering \citep{fau18,kow19} but not to explicitly infer  $K$ and the change points locations.

In this paper, we study spike and slab priors for offline multiple change point detection. Starting from a baseline $f_0$, we model each increment through the operator $\Delta f_i =f_i - f_{i-1}$ for $1\leq i \leq T$ and introduce latent binary variables $(Z_1,\ldots,Z_T)^{\top}$ to indicate whether  $\Delta f_i$ corresponds to a change point or not. The prior distribution on the increment $\Delta f_i$ under $Z_i=0$ is a distribution ``concentrated" around $0$ (or a point mass) called spike. The prior distribution  for $\Delta f_i$ under $Z_i=1$ is a diffused distribution called slab. The choice of which distributions to use for the spike, the slab, and the model space, has been a subject of extensive research; see \cite{bha19} for a recent review. We use the shrinking and diffusing prior of \cite{nar14}, which consists of Gaussian spike and slab priors with sample size dependent prior variances. The reasons for this choice are $(i)$ \cite{nar14} proved one of the strongest selection consistency results in the Bayesian variable selection literature, and $(ii)$ \cite{che19} recently proposed a methodology for fast Bayesian variable selection employing this prior and not requiring MCMC.

Here, we make the following contributions to the change point detection literature. $(i)$ We show how to employ spike and slab priors for change point detection and propose a fast algorithm that does not rely on MCMC for this task, making it one of the fastest Bayesian methods available. $(ii)$ We establish that a modified estimator based on the shrinking and diffusing prior is consistent and achieves optimal localization rates of multiple change points. We show that this optimality also holds when the estimator is based on our fast algorithm in the single change point detection framework. $(iii)$ Through simulations, we show that our procedure is competitive with state-of-the-art methodologies. A salient feature of our approach is that it is highly robust to misspecification of the noise term, a situation where many state-of-the-art benchmarks fail by substantially overestimating the number of change points. 

While in this paper we study the univariate change point detection problem, methods for  change point detection have also been studied for other types of data beyond univariate mean change point detection and settings more general than \eqref{eq:model}.   \cite{pei17}  considered  change point detection with heterogeneous noise.  \cite{carlstein1988nonparametric,rizzo2010disco,zou2014nonparametric,matteson2014nonparametric,padilla2018sequential,padilla2019optimal,padilla2019optimal2} developed nonparametric change point methods that can detect arbitrary changes in distribution. \cite{cho2015multiple,cho2016change,wang2016high} focused on  high-dimensional change point estimators.  \cite{aue2009break,avanesov2018change,wang2017optimal} studied covariance change point detection.  \cite{fearnhead2018changepoint} considered methods for change point detection combining a robust loss with the $\ell_0$ penalty. \cite{vanegas2019multiscale} proposed a multiscale method for quantile change point detection. Here, we focused on a simpler setting because spike and slab priors were not studied in the change point detection context, and, as stated by \cite{wang2020univariate}, the estimators built for \eqref{eq:model} are often the building blocks for more complex settings. 

% \cite{padilla2019change3,yu2021optimal} studied change point detection when the observations are binary networks.

The rest of the paper is organized as follows. Section \ref{sec:mod} describes the model, conditions on the prior parameters, and introduces the fast algorithm. Section \ref{sec:theory} presents our main results on consistency. We present simulation studies in Section \ref{sec:sim} to illustrate how our proposal fares with existing procedures. Section \ref{sec:app} includes applications to microarray and ion channel data. Section \ref{sec:disc} concludes with a discussion on the use of spike and slab priors for multiple change point detection.

\section{Method}
\label{sec:mod}

We assume that $(\epsilon_t)_{1:T}$ are \iid Gaussian random variables with mean zero and known variance $\sigma^2$.  Let $\mathcal{C}^*:=\{\eta_0,\ldots, \eta_K\}$ denote the set of change points of the piecewise-constant right continuous signal $f_t$, % with $0=\eta_0 < \eta_1 < \ldots < \eta_K < T$. The signal is such that $f_{\eta_i} = \ldots = f_{\eta_{i+1}-1} \neq f_{\eta_{i+1}}$.
we can write this as 
\begin{equation}
	\label{eqn:model}
	f_t \,=\, \begin{cases}
		\mu_0   &   \text{if }\,\,\,\,  t  \leq \eta_1-1\\
		\mu_1   &   \text{if }\,\,\,\,  \eta_1  \leq t  \leq \eta_2-1\\
		\vdots\\
		\mu_K & \text{if }   \,\,\,\,  \eta_K \leq t  \leq T,\\
	\end{cases}
\end{equation} 
Further technical conditions on $f_t$ will be given in Section \ref{sec:theory}. A working model employing spike and slab priors is as follows:
\begin{equation}
\label{eq:basad.cp}
\begin{array}{lll}
Y_t|f_t, \sigma^2 &\sim& \mathcal{N}(f_t,\sigma^2)  \\
\Delta f_t| \sigma^2, Z_t=0, \tau^2_{0,T} &\sim& \mathcal{N}(0, \sigma^2 \tau^2_{0,T}),\\
\Delta f_t | \sigma^2, Z_t=1, \tau^2_{1,T} &\sim&  \mathcal{N}(0, \sigma^2 \tau^2_{1,T}), \\
P(Z_t=1)&=&1-P(Z_t=0)=q_T, 
\end{array}
\end{equation}
for $t$ from $1$ to $T$, $\Delta f_t = f_t -f_{t-1}$, $\Delta f_1 = f_1$, and $\tau^2_{0,T},\tau^2_{1,T}$, and  $q_T$ are hyperparameters that depend on $T$, with $\tau^2_{1,T}\gg \tau^2_{0,T}$. The rationale behind this set-up is that the posterior probability of $Z_{t}=1$ should be high for $t\in \mathcal{C}^*$; vice versa, the posterior probability of $Z_t=0$ should be high for $t \notin \mathcal{C}^*$. A natural change point detection procedure is to employ  the posterior probabilities of $Z_t$ to determine if $t$ is a change point or not, for example classifying $t$ as a change point if the posterior exceeds a certain threshold. We elaborate on this selection rule later.

 A relevant difference with variable selection comes from the fact that while covariates are not ordered, in change point detection we generally want to avoid classifying consecutive time instances as change points. We expect this behavior because it is a common feature in the change point detection literature. Most of the procedures employ minimum spacing conditions, \textit{i.e.} the distance between consecutive change points is lower bounded by a quantity $\Delta >0$, such that $|\eta_j-\eta_{j+1}|>\Delta$. Minimum spacing conditions are used both in the finite sample implementations of the estimators and in the proofs of consistency. We will introduce a procedure to avoid consecutive change points.   %We assume that  $t$ is a change point if $P(Z_t=1| \textbf{Y},\sigma^2)>0.5$. 

The model \eqref{eq:basad.cp}, which we will refer to as \textit{basad.cp}, is the analogue of the \textit{basad} variable selection procedure of \cite{nar14} to change point detection. The sample size dependent hyperparameters $\tau^2_{0,T},\tau^2_{1,T}$ and $q_T$, are the salient feature of \eqref{eq:basad.cp}. We require that as $T\to \infty$,  $\tau^2_{0,T} \to 0$ and $\tau^2_{1,T} \to \infty$. In variable selection, a shrinking $\tau^2_{0,T}$ ensures that the marginal posterior probability of including (excluding) an active (inactive) covariate converges to one as sample size increases.  Increasing $\tau^2_{1,T}$ and $q_T$ allows for the consistent estimation of the number of active covariates and consistent model selection. \cite{nar14} proved that the penalization achieved through $\tau^2_{1,T}$ and $q_T$ is  equivalent to an explicit $\ell_0$ penalty. We will show in Section \ref{sec:theory} that these parameters play a similar role for achieving consistent estimation of $K$ and $\eta_1,\ldots,\eta_K$. Note that a similar asymptotic result holds despite the settings being very different: in high-dimensional variable selection, the number of covariates grows at a rate faster than the number of samples; in change point detection, the number of piecewise increments is equal or smaller than the sample size. 

Another key feature of the methodology of \cite{nar14} is that they employ the marginal posterior probabilities $P(Z_t=1| \textbf{Y},\sigma^2)$ to select the active parameters in the finite sample implementation (to prove consistency they employ $P(\textbf{Z}| \textbf{Y},\sigma^2)$, with $\textbf{Z}=(Z_1,\dots, Z_T)^{\top}$). The idea is motivated by computational reasons, given that one can sample from the marginals with a Gibbs sampler, which is not available for $P(\textbf{Z}| \textbf{Y},\sigma^2)$. Furthermore, the MCMC targeting $P(\textbf{Z}| \textbf{Y},\sigma^2)$ has a much larger space of models to explore. MCMC employed in Bayesian change point detection also struggle to explore the state-space, \textit{e.g.} \cite{fea06,rig12}. For this reason, we will follow the same approach for change point detection. %The recent work by \cite{liu20} also employs marginal posterior probability for model selection. 

We classify a time instance $t$ as a change point if $P(Z_t=1| \textbf{Y},\sigma^2)$ is larger than a certain threshold. In this case, the estimated number of change points $\widehat{K}$ is the number of marginal posterior probabilities larger than the chosen threshold. The model selected using $0.5$ as a threshold corresponds to the median probability model of \cite{bar04}, who also proved that it is the optimal predictive model. An alternative strategy to select the change points would be to first rank the $\Delta f_t$ based on $P(Z_t=1| \textbf{Y},\sigma^2)$, and then select the top $\widehat{K}$ (the model size) increments according to a given information criteria. We do not investigate this strategy and leave it for future work. We further stress that, under this selection rule, it is likely that there will be consecutive time instances such that $P(Z_j=1| \textbf{Y},\sigma^2)>0.5$, \textit{i.e} consecutive points could be ``classified" as change points. We deal with this issue after introducing an alternative methodology to compute $P(Z_t=1| \textbf{Y},\sigma^2)$.

Recently, \cite{che19} introduced a sequential procedure based on a misspecification of \textit{basad} that admits marginal posterior probabilities in closed form. Their method, called \textit{solo spike and slab}, has asymptotic properties and empirical accuracy similar to \textit{basad}, while being substantially faster. While our setting can be seen as a particular instance of the linear regression framework of \cite{che19}, there is an advantage in deriving the closed form marginal $P(Z_t=1| \textbf{Y},\sigma^2)$ for the setting considered in this paper because we will be able to simplify certain calculations further. We do this following the same steps of \cite{che19}. Suppose we are interested in testing whether a time instance $j$ is a change point,  we could consider the following model:
\begin{equation}
\label{eq:solo.cp}
\begin{array}{lll}
Y_t|f_t, \sigma^2 &\sim& \mathcal{N}(f_t,\sigma^2), \,\,\,\,\,\,\, t =1 \,\ldots,T,  \\
\Delta f_j| \sigma^2, Z_j=0, \tau^2_{0,T} &\sim& \mathcal{N}(0, \sigma^2 \tau^2_{0,T}), \\
\Delta f_j| \sigma^2, Z_j=1, \tau^2_{1,T} &\sim& \mathcal{N}(0, \sigma^2 \tau^2_{1,T}),\\
\Delta f_{i} | \sigma^2, \tau^2_{T} &\sim& \mathcal{N}(0, \sigma^2 \tau^2_{T}),     \,\,\,\,\,\,\, i\in \{1 \,\ldots,T\}\backslash \{j\},    \\
%\Delta f_{-j} | \sigma^2, \tau^2_{T} &\sim& \mathcal{N}(0, \sigma^2 \tau^2_{T}),  \\
P(Z_j=1)&=& 1-P(Z_j=0)=q_T, 
\end{array}
\end{equation}
%where $-j$ takes value in $\{1, \ldots,j-1,j+1,\ldots,T\}$, and 
where  $\tau^2_{T}$ is an additional sample size dependent hyperparameter. Thus,  we place a spike and slab prior on a single change point at a time (in this case $\Delta f_j$), and place conjugate Gaussian priors on the remaining terms. The tuning parameter $\tau^2_{T}$ controls  the shrinkage across time instances. The advantage of model \eqref{eq:solo.cp} is that it allows us to write the marginal posterior probabilities in closed-form. First, we can marginalize out $\bm{\Delta f_{-j}}:=(\Delta f_1, \ldots,\Delta f_{j-1},\Delta f_{j+1},\ldots,\Delta f_T)^{\top}$ in the likelihood $\mathcal{L}(\textbf{Y}| \bm{\Delta f_{-j}}, \Delta f_j, \sigma^2)$ to compute the marginal likelihood
\begin{equation}
\label{eq:marglik}
\mathcal{L}(\textbf{Y}|  \Delta f_j, \sigma^2) \propto \exp \{  \frac{ -\Delta f_j^2 (T-j -\sum_{j+1}^T {n'_{i}}) \gamma_{j,j} + 2\Delta f_j \overline{y''_{j, j}}}{2 \sigma ^2}\}.
\end{equation}
The parameters $n'_{j+1},\ldots, n'_{T}, \gamma_{j,j}$ and $\overline{y''_{j, j}}$  are computed recursively as follows: initialize $n'_T= \tau_T^2/(\tau_T^2+\sigma^2)$, and compute recursively for $i=T-1$ to $j+1$ 

\begin{align}
\label{eq:singn_para1}
n'_{i}=\frac{\tau_T^2 (T-i-1 - \sum_{k=i+1}^T n'_{k})^2}{\tau_T^2 (T-i-1 - \sum_{k=i+1}^T n'_{k})+\sigma^2},  \,\,\,\ \text{and} \,\,\,\ 
\overline{y'_{i}}=\frac{\sum_{t=i}^T y_{t} - \sum_{k=i+1}^T n'_{k} \overline{y'_{k}}}{T-i-1 - \sum_{k=i+1}^T n'_{k}}.
\end{align}
Then, set $\gamma_{j,1}=1$, and for $i=1$ to $j$ compute 
\begin{align}
\label{eq:singn_para2}
n''_{i,j}&=\frac{\tau_T^2}{\tau_T^2 \gamma_{i,j} (T-i-1 - \sum_{k=j+1}^T n'_{k})+\sigma^2}, \nonumber\\
\overline{y''_{i,j}}&=\sum_{t=i}^T y_{t} - \sum_{k=i+1}^T n'_{k} \overline{y'_{k}}- \Big(T-i-1- \sum_{k=j+1}^Tn'_{k}\Big)\sum_{k=1}^{i-1} n''_{k,j} \gamma_{k,j} \overline{y''_{k,j}}, \\
\gamma_{i+1,j}&=1-\Big(T-i-1- \sum_{k=j+1}^T n'_{k}\Big)\Big(\sum_{k=1}^{i} n''_{k,j} \gamma_{k,j}^2 \Big). \nonumber
\end{align}
Despite the involved notation, simple calculations lead to the definitions of parameters in \eqref{eq:singn_para1} and \eqref{eq:singn_para2}. The basic idea is first to marginalize $\Delta f_T$, then $\Delta f_{T-1}$, then continue sequentially to $\Delta f_{j+1}$. This first step leads to the definition of the parameters in \eqref{eq:singn_para1}. In the second step, we first marginalize $\Delta f_1$ and then recursively integrate out the remaining parameters until $\Delta f_{j-1}$. This second step leads to the definition of the parameters in \eqref{eq:singn_para2}.
 
Given the marginal likelihood $\mathcal{L}(\textbf{Y}|  \Delta f_j, \sigma^2)$, we can compute the marginal posterior distribution of $\Delta f_j$ through Bayes rule:
\begin{equation}
\label{eq:marg.post}
\Delta f_j |\textbf{Y}, \sigma^2, q_{T}, \tau^2_{0,T},\tau^2_{1,T} \sim (1-q_{T}) \omega_{0,j} N(\mu_{0,j},\xi_{0,j}^2) + q_{T}\omega_{1,j} N(\mu_{1,j}, \xi_{1,j}^2), 
\end{equation}
where the parameters  are defined as follows,  for $k \in \{0,1\}$:
\begin{align}
\label{eq:post.param}
\mu_{k,j}&=\frac{\overline{y''_{j,j}}}{(T-j-1 - \sum_{k=j+1}^Tn'_{k})\gamma_{j,j} +\tau_{k,T}^{-2} \sigma^2}, \nonumber\\
\xi_{k,j}&=\frac{\sigma^2}{(T-j-1 - \xi_{k=j+1}^Tn'_{k})\gamma_{j,j} +\tau_{k,T}^{-2} \sigma^2}, \\
\omega_{0,j} &=\exp\Bigg\{\frac{1}{2 \sigma^2} \frac{\overline{y_{j,j}^{'' 2}}}{(T-j-1 - \sum_{k=j+1}^T n'_{k})\gamma_{j,j} +\tau_{k,T}^{-2} \sigma^2} 
 \Bigg\} \sqrt{\frac{\tau_{k,T}^{-2}}{(T-j-1 - \sum_{k=j+1}^T n'_{k})\gamma_{j,j} +\tau_{k,T}^{-2} \sigma^2}}.\nonumber
\end{align}
The parameters above are all we need to compute $P(Z_j=1| \textbf{Y},\sigma^2)$, which corresponds to 
\begin{equation}
\label{eq:test}
P(Z_j=1| \textbf{Y},\sigma^2)= \frac{q_{T} \omega_{1,j}}{q_{T} \omega_{1,j}+(1-q_{T}) \omega_{0,j}}.
\end{equation}
Given \eqref{eq:test}, we follow the same procedure described for \textit{basad.cp}: a time instance is declared a change point if $P(Z_j=1| \textbf{Y},\sigma^2)$ exceeds the prespecified threshold. In practice, we would not be interested only in a single time instance $j$, so one needs to compute \eqref{eq:test} for $j$ in $1$ to $T$. \textit{i.e} we are considering $T$ models. 

 Note that the parameters in \eqref{eq:singn_para2} depend on a given time instance $j$, whereas the parameters in \eqref{eq:singn_para1} are shared across multiple points. The dependence of parameters \eqref{eq:singn_para2} on a given time instance $j$ represents the main computational bottleneck of the \textit{solo.cp} algorithm, because they need to be recomputed $T$ times. The shrinkage effect of $\tau_T^2$ is explicit in all the $n_i'$'s and $n''_{i,j}$'s: samples that are closer to  time instance $j$ have a ``higher weight" in $\overline{y'_{i}}$ and $\overline{y''_{i,j}}$; on the other hand, the further we are moving away from $j$, the less informative the observations are. Sparsity is instead induced by $q_T$ in \eqref{eq:test}.

Equations \eqref{eq:marglik}, \eqref{eq:marg.post}, and \eqref{eq:test} are the analogues to $(7), (11)$ and $(19)$ in \cite{che19}. Similarly the definitions of parameters in \eqref{eq:post.param} are the analogues of $(12)-(14)$ from \cite{che19}. The differences arise because their definitions rely on a matrix of covariates and require several matrix multiplications and inversions. Importantly here, we can write analytically all the formulas and bypass the need for these matrix operations.

%We will study theoretical properties of \eqref{eq:basad.cp} and \eqref{eq:solo.cp} in Section \ref{sec:theory}. For finite sample sizes, we note that there could be consecutive time points such that $P(Z_j=1| \textbf{Y},\sigma^2)>0.5$, \textit{i.e} consecutive points could be ``classified" as change points. We expect this behavior because it is a common feature in the change point detection literature. Most of the procedures employ minimum spacing conditions, \textit{i.e.} the distance between consecutive change points is lower bounded $|\eta_j-\eta_{j+1}|>\Delta$, both in the finite sample implementations of the estimators and in the proofs of consistency. The \textit{solo.cp} procedure is particularly sensitive to this issue because we are allowing only one point at a time to be a change point and shrink everything else. 

Regardless of whether we compute $P(Z_{\eta_i}=1| \textbf{Y},\sigma^2)$ through \textit{basad.cp} or \textit{solo.cp}, we propose the use of a post-processing step to avoid the detection of consecutive change points. This involves a rule that defines when two or more estimates will be considered as ``consecutive", and a selection rule to determine which estimates to keep. 

 In detail, let $\widehat{\mathcal{C}_0}:=\{\widehat{\eta_1},\ldots, \widehat{\eta}_{\widehat{K}_0}  \}$ be the set of points such that $P(Z_{i}=1| \textbf{Y},\sigma^2)>0.5$ for $i$ in $1$ to $T$. Now, fix $\Delta\in \mathbb{N}$, and partition $\widehat{\mathcal{C}_0}$ into nonempty subsets $\widehat{\mathcal{C}_0}^1,\ldots, \widehat{\mathcal{C}_0}^{\widehat{K}}$ such that for all $\eta \in \widehat{\mathcal{C}_0}^i$ there exists at least one $\eta' \in \widehat{\mathcal{C}_0}^i$ (if $\widehat{\mathcal{C}_0}^i \backslash\{\eta\} \neq \emptyset $) such that $|\eta-\eta'|\leq \Delta$ and no $\eta'' \in \widehat{\mathcal{C}_0} \setminus \widehat{\mathcal{C}_0}^i$ such that $|\eta-\eta''|\leq \Delta$. Hence, the partition defines the notion of  ``consecutive change points".  Finally, within each subset $\widehat{\mathcal{C}_0}^i$, choose the point $\eta_i= \underset{\eta \in \widehat{\mathcal{C}_0}^i}{\arg \max} P(Z_{\eta}=1| \textbf{Y},\sigma^2) $. The estimated set of change points is $ \widehat{\mathcal{C}}:=\{\eta_1,\ldots, \eta_{\widehat{K}}\}$. 

	\begin{algorithm}[!t]
	\caption{Spike and slab change point detection}
	\label{all:all}
	\begin{algorithmic}
		\State \textbf{Inputs:} \textbf{Y}, T, $\Delta$
		\State \textbf{Output:} $\widehat{\mathcal{C}}$, $\widehat{K}$, $\sigma$ 
		\begin{enumerate}
			%	\item Define $c_n^v$ for all $v \in V$
			\item Compute $P(Z_1=1| \textbf{Y},\sigma^2) ,\ldots, P(Z_T=1| \textbf{Y},\sigma^2)$ 
			
			\textbf{If} \textit{basad.cp}
			\begin{itemize}
				\item Approximate $P(Z_1=1| \textbf{Y},\sigma^2) ,\ldots, P(Z_T=1| \textbf{Y},\sigma^2)$ with the Gibbs sampler defined in \cite{nar14}.
			\end{itemize}
		
		\textbf{If} \textit{solo.cp} \textbf{for} $i-1,\ldots,T$ \textbf{do}
		\begin{itemize}
			\item Compute posterior parameters \eqref{eq:post.param}
			\item Compute  $P(Z_i=1| \textbf{Y},\sigma^2)$ through \eqref{eq:test}.
		\end{itemize}
			\item Define $\widehat{\mathcal{C}_0}:=\{i: P(Z_i=1| \textbf{Y},\sigma^2)>0.5\}$
			\item Partition $\widehat{\mathcal{C}_0}$ into subsets of nonconsecutive change points $\widehat{\mathcal{C}_0}^1,\ldots, \widehat{\mathcal{C}_0}^{\widehat{K}}$ (see the main text)
			\item Set  $ \widehat{\mathcal{C}}:=\{ \eta_i= \underset{\eta \in \widehat{\mathcal{C}_0}^i}{\arg \max} P(Z_{\eta}=1| \textbf{Y},\sigma^2), \text{for } i=1,\ldots,\widehat{K}  \}$
		\end{enumerate}
	\end{algorithmic}
\end{algorithm}

A few remarks. First, the length of the partition determines the number of estimated change points $\widehat{K}$. %Second, $P(Z_{\eta_i}=1| \textbf{Y},\sigma^2)$ can be computed either with \textit{basad.cp} or \textit{solo.cp}. Hence, the post-processing procedure can be used with the output of either method. 
Second, we need an extra parameter $\Delta$ to define the partition of $\widehat{\mathcal{C}_0}$. The sensitivity of the two methods to $\Delta$ is studied in Section \ref{sec:sim}. Lastly, we pick the time instance having the maximum marginal posterior probability within each subset and classify it as the change point. Whereas this is an arbitrary criterion, choosing the point that maximizes a given test statistics is  standard in the change point detection  literature; see for instance  \cite{fryzlewicz2014wild}. 

 Algorithm \ref{all:all} summarizes the methodology. An input of the algorithm is $\sigma$, which we assumed to be known. In many applications this assumption does not hold and we require an estimate. For example, the \textit{wbs}  \citep{fryzlewicz2014wild} and \textit{r-fpop} \citep{fearnhead2018changepoint} employ the median absolute deviation estimator. The \textit{ebpiece} method of \cite{liu20} uses the fused LASSO residuals standard deviation computed through the ``one standard error" rule \citep{tib05}.

In this section, we assumed that the number of observations collected at a given point $t$ ($n_t$) is equal to one (case $n_t=1$), which is the standard in the literature. In applications, this may not be the case (case $n_t>1$). This situation could arise if multiple observations are collected at once, or if observations are collected at distinct time points, but the reported data are binned into time intervals. To our knowledge, there are few methods in the literature dealing with this situation \citep{padilla2019optimal}.  The extension of \textit{basad.cp} and \textit{solo.cp} to the case $n_t>1$ is straightforward.  Parameters \eqref{eq:singn_para1}, \eqref{eq:singn_para2}, and \eqref{eq:post.param} can be written in closed form, including an adjustment done through $n_t$ for all $t$.  The explicit formulas are provided in Appendix \ref{app:solo.cpnt>1}. In Section \ref{sec:sim}, we show that the case $n_t>1$ is particularly beneficial in terms of empirical performance for the two methodologies discussed in this section.

\section{Theory}
\label{sec:theory}

In  this section, we provide some theoretical support for the methods that we study in this paper. Our theory is organized into subsections. First, we show that for the task of multiple change point detection, a modified version of the estimator described in \eqref{eq:basad.cp} (\textit{basad.cp}) based on the spike and slab variable selection framework of \cite{nar14} leads to optimal localization rates of the change points. Specifically, for the case of a bounded number of change points, under a slightly weaker signal-to-noise condition than the wild binary segmentation and $\ell_0$ based methods, we attain optimal localization rates. We then show that this optimality is also preserved in the single change point detection framework  if we consider a version of the \textit{solo.cp} estimator. For this section, we ignore the post selection step described in Section \ref{sec:mod}.

\subsection{Multiple  change point detection with shrinking and diffusion priors}
\label{sec:th1}
%5	 SHRINKING AND DIFFUSING PRIORS}

%\sout{Consider  data  generated  from the  model}
%\begin{equation}%
%	\label{eqn1}
%	Y_t \,=\, f_t^* + \epsilon_t, \,\,\,\,\,\,t=1,\ldots,T, \gr{[OUT]}
%\end{equation}
%\lc{where  $\epsilon \sim N(0,\sigma^2 I_T)$  with  $I_T \in \mathbb{R}^{T\times T}$,
%and let  $\mathcal{C}^* = \{ j\,\,:\,   f_j^* \neq f_{j-1}^* \}$ be the set of jumps in the mean of the sequence vector $\textbf{Y} \in \mathbb{R}^T$.  Our goal  is to estimate  $\mathcal{C}^*$.}{[introduced in Sec 1 and 2]} 
We consider a modified version of the \textit{basad.cp} estimator defined as follows.  Let $m \in \mathbb{N}$  with  $m \leq  T$  and consider $\Lambda_1,\ldots,\Lambda_m$  a partition of  $\{1,\ldots,T\}$ such that  $\vert \Lambda_j \vert    =   T/m$ for all $j \in \{1,\ldots,m\}$. Let    $\widetilde{\textbf{Y}} \in \mathbb{R}^m$ be the statistic

\[
\displaystyle    \widetilde{Y}_j  \,=\, \frac{1}{   \sqrt{ \vert  \Lambda_j \vert  }  } \underset{i \in \Lambda_j  }{\sum} Y_i,
\]
for $j=1,\ldots,m$. 
We also define \[
	\displaystyle    \tilde{f}_j  \,=\, \frac{1}{   \sqrt{ \vert  \Lambda_j \vert  }  } \underset{i \in \Lambda_j  }{\sum} f_i,
	\]
	for $j=1,\ldots,m$. 
%\gr{[We are missing a definition of $\tilde{f}$, if there is no change point is clear, if there is a change point?]} 
It is convenient to rewrite \eqref{eq:basad.cp} for the data vector  $\widetilde{\textbf{Y}}$ as
\begin{equation}
	\label{eqn:bayesian}
	\begin{array}{lll}
		\widetilde{\textbf{Y}} \,|\,   \tilde{f}, \sigma^2   & \sim &     \mathcal{N}\left(    \tilde{f}, \sigma^2 I_{m}  \right),\\
		\Delta \tilde{f}_j\,|\,   \sigma^2, Z_{j} = 0 ,\tau_{0,m}^2   &\sim &     \mathcal{N}\left(   0, \sigma^2   \tau_{0,m}^2 \right),\\
		\Delta \tilde{f}_j \,|\, \sigma^2, Z_{j} = 1,\tau_{1,m}^2   &\sim &     \mathcal{N}\left(   0, \sigma^2   \tau_{1,m}^2 \right),\\
		P(Z_j= 1)    &\, =\,&   1- P(Z_j = 0) =  q_m,\,\,\,\,\,\,\, j =1 \,\ldots,m,\\
		%	\sigma^2   & \sim&   IG(\alpha_1,\alpha_2),
	\end{array}      
\end{equation}
where  $q_m ,\tau_{0,m},\tau_{1,m}>0$.  Furthermore, $\Delta \tilde{f}_1= \tilde{f}_1$, and  $\Delta \tilde{f}_j =  \tilde{f}_j - \tilde{f}_{j-1}$ for  $j = 2,\ldots, m$.
%Notice that $\beta_j$ represents a parameter for the jump size at location $j$. \gr{It is equivalent to the piecewise constant change $\Delta f_j$ used in the previous section. }% Thus for all 

The goal is to define an estimator $\widehat{\mathcal{C}}\subset \{1,\ldots,T\}$ of  $\mathcal{C}^*$. We do this by first defining an estimator $\widetilde{\mathcal{C}}$ relying on the indexes of the partition $1,\ldots,m$, then we use $\widetilde{\mathcal{C}}$ to construct our actual estimator. First, let  
\[
%\widetilde{\mathcal{C}}  
      \widetilde{ \textbf{Z} }   \,=\,     \underset{    \textbf{Z}    \in \{0,1\}^{m}  }{\arg \max}\,P\left(    \textbf{Z}     \,|\,  \widetilde{\textbf{Y}} ,  \sigma^2 \right)
\]
%\mathcal{C} \subset \{1,\ldots,m\}
%$ \{  j\,:\,  Z_j =1  \} =  \mathcal{C}$
and  $\widetilde{\mathcal{C}}     \,=\,\{ j \, :\, \widetilde{Z}_j =1  \} $. The set $\widetilde{\mathcal{C}}$ is then used to construct $\widehat{\mathcal{C}} \subset \{1,\ldots,T\}$ as follows:
\begin{itemize}
	\item  If  $i \in \widehat{\mathcal{C}}$  then  there exists  a  $j \in \widetilde{\mathcal{C}}$  with  $i \in \Lambda_j$.
	\item  If  $j \in  \widetilde{\mathcal{C}}$  then  for a unique   $i \in \widehat{\mathcal{C}}$  we have that  $i \in \Lambda_j$.
\end{itemize} 

Note that $\widetilde{\mathcal{C}}$ is constructed using the posterior distribution of \textbf{Z} rather than the marginals $P(Z_j=1| \textbf{Y},\sigma^2)$ (as discussed in Section \ref{sec:mod}). Furthermore, we are conditioning on $\widetilde{\textbf{Y}}$. The construction of $\widehat{\mathcal{C}}$ is used to map the estimates conditioned on the transformed data to the actual time indices we are trying to infer. Our results show that  $\widehat{\mathcal{C}}$   defined by the modified estimator based on $\widetilde{\textbf{Y}}$ and $\widetilde{\mathcal{C}}$ attains optimal  localization rates for estimating  $\mathcal{C}^*$.  Our result exploits Theorem 4.1 in \cite{nar14} which provides a consistency result for linear model estimation with the shrinking and diffusing prior. Our main result is based on the following modeling assumption.

\begin{assumption}
	\label{as1}
	Let  $\kappa$  be the minimum jump size, thus,  
	\[
	\kappa \,:=\,  \underset{j  \in  \mathcal{C}^* }{\min}     \,\vert  f_{j} - f_{j-1} \vert.
	\]
	Then we require that 
	\[
	\frac{  \kappa^2   T}{\sigma^2 \log T}   \,\rightarrow \,   \infty,
	\]
	as  $T \rightarrow \infty$.  Furthermore,  we impose the following minimum spacing condition
	\[
	\Delta  \,:=\,\underset{j   \neq  j^{\prime},   \,  j,j^{\prime} \in \mathcal{C}^*  \,\,\,\, }{\min}\,\vert j-j^{\prime} \vert\,\geq \, \frac{c_1   \sigma^2 \log T }{\kappa^2},
	\]
	for a large enough  $c_1>0$, 	and  require that  $K:=    \vert  \mathcal{C}^*\vert     =     O(1)$.
	
\end{assumption}

Assumption \ref{as1} can be thought as a signal-to-noise-ratio condition. In fact, Assumption \ref{as1} is a weaker condition than Assumption 2 from \cite{wang2020univariate} which states that
\[
\Delta  \,\geq \,  \frac{  c\sigma^2  \log^{1+\xi} T }{\kappa^2},
\]
for positive constants  $c$ and $\xi$. However, the framework in \cite{wang2020univariate}  allows the possibility that $K$ diverges whereas here we require that $K = O(1)$.  

We are now ready to state the main result of this section.

\begin{theorem}
	\label{thm1}
	Suppose that  Assumption \ref{as1} holds. Then  for a constant $c_0>0$ the estimator  $\widehat{\mathcal{C}}$ satisfies
	\[
	P\left(  \vert   \widehat{\mathcal{C}} \vert =  K,   \,\,\,\,  \,  \underset{\eta\in \mathcal{C}^*}{\max}  \,\,\underset{\hat{\eta} \in  \widehat{\mathcal{C}} }{\min}\,\vert \hat{\eta}-\eta\vert \,\leq\, \frac{c_0 \sigma^2 \log T }{\kappa^2}     \right)\,\rightarrow \,1,
	\]
	as  $T\,\rightarrow \,\infty$, provided that $\tau_{0,m}^2= o(1/m)$, $q_m  \asymp  1/m$ , and $\tau_{1,m}^2  \asymp      m^{1+3\delta}$ for some  $\delta>1$, and $m$ such that
		\[
	m \,\asymp\, \frac{\kappa^2   T }{\sigma^2\log T}.
	\]
	% $\tau_{0,m} <1/(m\sqrt{2})$.
\end{theorem}

Notably, Theorem  \ref{thm1} shows that the maximum a posteriori estimator constructed based on the model (\ref{eqn:bayesian}) attains a localization rate of order $\log T/\kappa^2$.  As \cite{wang2020univariate} showed, this localization rate is minimax optimal up to a logarithm factor.  Importantly, our guarantee on the localization rate holds under the minimum signal-to-noise ratio condition possible; see Lemmas 1--2  in \cite{wang2020univariate}.

%\lc{Unfortunately,  \textit{basad.cp}  requires a Gibbs sampler,  which means that computing  $\widetilde{\mathcal{C}}$ and  hence  $\widehat{\mathcal{C}}$ can be computationally intensive.}
We stress that in this section we considered the joint posterior probability of $\textbf{Z}=(Z_1,\ldots,Z_m)^{\top}$, while in Section \ref{sec:mod} we discussed the use of marginal posterior probabilities for finite sample implementation of \textit{basad.cp}. In practice we use the fast method described  in the previous section  based on a misspecification of \eqref{eqn:bayesian}. Next, we show that such surrogate procedure  still enjoys a localization guarantee in the case of single change point detection.

\subsection{Localization rate of the fast Bayesian  estimator in single change point setting}
\label{sec:fast}

Throughout this section we assume the  model described by (\ref{eqn:model}) but in the presence of only one change point, thus  $\vert \mathcal{C}^*\vert \,=\,1$.  Under such setting, we study the behavior of the posterior means involved in  \textit{solo.cp}, the fast detection procedure proposed in Section \ref{sec:mod}. Notably, while our estimator is a particular instance of the high-dimensional linear framework from \cite{che19}, the theory from \cite{che19} cannot be directly applied in our setting. The reason is that when writing (\ref{eqn:model}) as a linear model 
the design matrix does not satisfy the conditions required for consistency in \cite{che19}. Despite this, we show that a version of  our  fast estimator attains optimal localization rates for single change point detection.

%With the notation  from Section \ref{sec:mod}, we now show \red{that a change point selection criterion based on the posterior means of the \textit{solo.cp} method}, lead to consistent change point detection. 
Throughout the section we consider the following change point selection criterion
\[
\hat{j}  \,:=\,\underset{ j:\,   \,    \min\{T-j,j\}\geq cT   }{  \argmax}\, \left\vert\frac{\mu_{1,j}  +  \mu_{1,T-j+1}^{\prime}}{2}\right\vert , 
\]
where $\{\mu_{1,j}\}$ is the vector of posterior means defined in \eqref{eq:post.param}, $\{\mu_{1,j}^{\prime}\}$ is the version of  $\{\mu_{1,j}\}$  based on the vector $(-Y_T,\ldots,-Y_1)^{\top}$  instead of  $(Y_1,\ldots,Y_T)^{\top}$, and $c>0$. The vector $\{\mu_{1,j}^{\prime}\}$ is employed to obtain the desired localization rate. The need for this second vector will become apparent in the proof of Theorem \ref{thm2}.
	
The criterion $\hat{j} $ has several notable differences with \textit{solo.cp}: $(i)$ it ignores the spike components, $(ii)$ it does not use the posterior marginals of the $Z_j$s to select the change point, $(iii)$ it requires to compute $2T$ posterior means, using the data set twice. However, we deem important to study the behavior of this second estimator because there are important similarities between the two: $(j)$ they are based on the same model, $(jj)$ they employ the same idea of testing one change point at a time, $(jjj)$ they involve sample size dependent hyperparameters, $(jjjj)$ they employ the same posterior means.

\begin{theorem}
	\label{thm2}
	Let $\mathcal{C}^* =\{       j_0 \,:\,  f_{j_0} \neq f_{j_0-1} ,\,\,\,j_0>1   \}  $ and suppose that $|\mathcal{C}^*|=1$. In addition assume that:
	\begin{itemize}
		\item There exists a constant $c>0$  such that $\min\{      j_0, T-j_0  \} \geq c T$.
		\item The   sequence $\tau_T$   converges to  zero fast enough. 
		%\gr{[shall we have a rate here?]}
		\item The parameter  $\tau_{1,T}$ satisfies $\tau_{1,T }^{2}   \gtrsim T^{-1}$.
		%\[
		%  \frac{\max\{   \tau_{1,n}^2 n,1 \}}{  \vert  \beta_{j_0}^* \vert } \,=\,O(1),
		%\]
		\item   The jump size     $\kappa\,:=\, \vert f_{j_0}   -  f_{j_0-1}  \vert$ satisfies  $\kappa\gtrsim  \sigma \sqrt{\log T/T} $.
		
		Then there exists a constant  $c_1>0$ such that,   with probability approaching one, we have that
		\[
		\underset{j\,:\,   \,    \min\{T-j,j\}\geq cT,     \,\,       \vert  j-j_0\vert \geq c_1 \sigma^2\log T/   \kappa^2 }{\max} \, \left \vert\frac{ \mu_{1,j} +\mu_{1,T-j+1}^{\prime} }{2}\right\vert \,<\, \left \vert \frac{\mu_{1,j_0} +\mu_{1,T-j_0+1}^{\prime}}{2}\right\vert.
		\]
%	\sout{	where  $\{\mu_{1,j}^{\prime}\}$ is the version of  $\{\mu_{1,j}\}$  based on the vector $(-Y_T,\ldots,-Y_1)^{\top}$  instead of  $(Y_1,\ldots,Y_T)^{\top}$. }
	\end{itemize}
	
\end{theorem}

Theorem  \ref{thm2}  states that  in the single change point detection setting, if we detect the change point  based on the criterion $\hat{j}$, then we attain the localization rate $\sigma^2  \log T/\kappa^2$. Thus,
\[
\vert  \hat{j} - j_0 \vert \,\leq \, c_1\frac{\sigma^2 \log T}{\kappa^2},
\]
with probability approaching one. This localization rate is nearly optimal and matches the localization rate  from Theorem \ref{thm1}, and that $\ell_0$  regularization and wild binary segmentation also achieve (see  \cite{wang2020univariate}). Although our result here only allows one single change point,  the signal-to-noise ratio condition ($\kappa\gtrsim  \sigma \sqrt{\log T/T}$) in Theorem \ref{thm2}  is slightly weaker than those in previous work.

Another condition that is remarkably weaker is that on  $\tau_{1,T }^{2}$, which it is assumed $ \tau_{1,T }^{2}  \gtrsim T^{-1}$. In Theorem \ref{thm1}, we had $\tau_{1,T }^{2}$ diverging. The difference can be mostly explained by the fact that we do not use the spike components in this modified version of the \textit{solo.cp} method. This can be seen in the proof: since we are not using \eqref{eq:test} to select the change point, there is no need to assume $\tau_{1,T } \to \infty$ and $\tau_{0,T } \to \infty$.

%\gr{[I think we need a connection more clearly explained between estimator $\widehat{j}$ and the actual method we use, that is \eqref{eq:test} greater than a threshold. It looks to that this is direct, but it is worst commenting on]}

Notice that the fact  the change point estimation criterion $\hat{j}$ leads to consistent estimation does not necessarily imply  that the \textit{solo.cp} estimator attains optimal rates. 	 However, it shows that  an estimator closely related to the \textit{solo.cp} estimator has a desirable property, and our experiments on both real and simulated data will confirm excellent performance of \textit{solo.cp} as described in Section \ref{sec:mod}.

\section{Simulations}
\label{sec:sim}

We rely on simulations to explore the ability of the \textit{solo.cp} and \textit{basad.cp} estimators to accurately estimate $K$ and change point locations $\eta_1,\ldots, \eta_K$.  We consider realistic scenarios designed to capture the variability encountered in applications, varying the conditional mean $f_t$ and the distribution of the error terms $(\epsilon_t)_{1:T}$. We compare \textit{basad.cp} and \textit{solo.cp} with several state-of-the-art methods: \textit{wbs} \citep{fryzlewicz2014wild}, \textit{ebpiece} \citep{liu20}, \textit{smuce} \citep{fric14}, \textit{pelt} \citep{killick2012optimal}, and \textit{r-fpop} \citep{fearnhead2018changepoint}. We employ default settings in the implementations of these methods. Details are given in Appendix \ref{app:sim.det}. All code to reproduce the results in this section is available at \texttt{https://github.com/lorenzocapp/solocp\_experiments}. The methodology is available as a \texttt{R} package available for download at \texttt{https://github.com/lorenzocapp/solocp}

Our  empirical comparisons assess the accuracy of the different estimators with the following criteria.  We consider the statistic $K-\widehat{K}$ to measure how well each estimator recovers the true number of change points. We consider an order-invariant Haussdorf metric $d(\widehat{\mathcal{C}},\mathcal{C}^*)= d(\widehat{\mathcal{C}}|\mathcal{C}^*)+d(\mathcal{C}^*|\widehat{\mathcal{C}})$, where 
	$d(\widehat{\mathcal{C}^*}|\mathcal{C}^*)=\underset{\eta \in \mathcal{C}}{\max} \,\ \underset{x \in \widehat{\mathcal{C}}}{\min} | x- \eta|$ and $d(\mathcal{C}^*|\widehat{\mathcal{C}})=\underset{\eta \in \widehat{\mathcal{C}}}{\max} \,\ \underset{x \in \mathcal{C}^*}{\min} | x- \eta|$ are respectively the one-sided Haussdorf distances.  We use $d(\widehat{\mathcal{C}},\mathcal{C}^*)$ to assess the overall accuracy of the estimators in recovering the true change points locations $\eta_1,\ldots, \eta_K$. We employ $d(\widehat{\mathcal{C}},\mathcal{C}^*)$ in lieu of $d(\widehat{\mathcal{C}}|\mathcal{C}^*)$, being the latter insensitive to overestimation. Lastly, for all $\eta \in \mathcal{C}^*$ we calculate $\underset{x \in \widehat{\mathcal{C}}}{\min} | x- \eta|$, and report the proportion of points that are at distance zero, one, two, and equal or greater than three. We refer to this criterion as the normalized empirical distribution and denote it by $|\widehat{\eta}-\eta|/K$. It is a finer measure than the Haussdorf distance of the change point location estimation accuracy. Since this criterion is also insensitive to overestimation, we include the reciprocal $|\eta-\widehat{\eta}|/\widehat{K}$. The unnormalized version of this criterion is also considered by \cite{fryzlewicz2014wild}. 
	
\begin{figure}[!t]
	\begin{center}
		\includegraphics[scale=0.60]{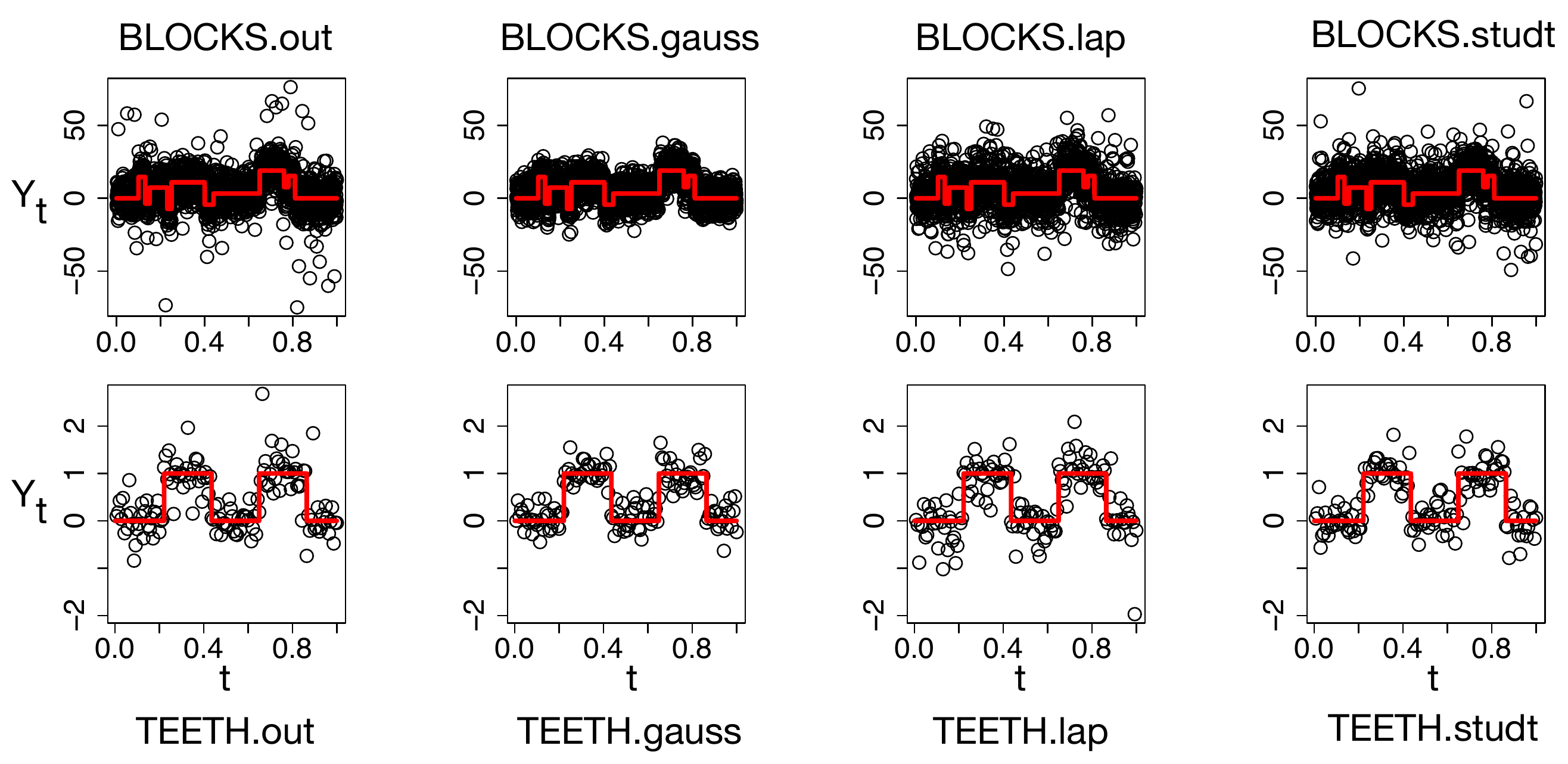}
	\end{center}
	\caption{\small{\textbf{Case $n_t=1$: examples of datasets for the eight scenarios considered and true test signals.} First row panels depict sample datasets along with the BLOCKS test signal (red), second row panels sample datasets along with the TEETH test signal (red). First column panels depict observations sampled with mixture of Gaussian errors (.out), the second column panels with the Gaussian errors (.gauss), the third column panels with Laplace errors (.lap), and the fourth column with Student's t errors (.studt).}}
	\label{fig:sim1}
\end{figure}

\subsection{Case $n_t=1$}
\label{sec:sim_nt1}
				
We consider two test signals and four error distributions. The first signal $f_t$ is called BLOCKS ($K=11, T=2048$), a standard benchmark for change point detection procedures (\textit{e.g.} used by \cite{fryzlewicz2014wild}), the second test signal is called TEETH ($K=4,T=140$). We consider four distributions for the error terms: Gaussian, Laplace, Student's $t$, and a mixture of Gaussians (to mimic the presence of outliers one of the two components has a larger variance). Change point locations of the test signals and the parameters of the error terms are fully specified in Appendix \ref{app:sim.det}. For each combination $(f_t , \epsilon_t)$, we sample $100$ datasets and report the average value for each criterion considered. Figure \ref{fig:sim1} plots examples of  data sets sampled for each scenario, along with the true test signals in red.

\begin{table}[!t] \centering 
		\renewcommand{\arraystretch}{0.85} % this reduces the vertical spacing between rows
		\caption{\small{\textbf{Case $n_t=1$:  Haudorff distance, empirical distributions and estimation bias in $K$ of the procedures considered for the BLOCKS test signals.} Average statistics computed over $100$ simulations for \textit{solo.cp}, \textit{ebpiece} \citep{liu20},  \textit{smuce} \citep{fric14},  \textit{wbs} \citep{fryzlewicz2014wild},  \textit{pelt} \citep{killick2012optimal}, and \textit{r-fpop} \citep{fearnhead2018changepoint}. In ``Data", .out refers to mixture of Gaussian errors, .gauss to Gaussian errors, .lap to Laplace errors, and .studt to Student's t errors. For $|\widehat{\eta}-\eta|/K$ and $|\eta-\widehat{\eta}|/\widehat{K}$ the higher the number in the zero column the better. Conversely, for $d(\widehat{\mathcal{C}},\mathcal{C}^*)$ the lower the better. For $K-\widehat{K}$, the closer to the zero the better. We report in bold the methods with best empirical performance and those within $10\%$ of the best. The method \textit{basad.cp} is  not included since it required a computing time longer than two hours. We include between brackets the computing time for \textit{ebpiece} if initialized with the in-built procedure (see text for a discussion).}}
	\label{tab:block} 
	\scalebox{0.8}{
		\hspace{-0.5cm}\begin{tabular}{@{\extracolsep{5pt}} lc|cccc|cccc|c|c|c} 
			\\[-1.8ex]\hline 
			 \\[-1.8ex] 
			& &\multicolumn{4}{c}{$|\widehat{\eta}-\eta|/K$} & \multicolumn{4}{c}{$|\eta-\widehat{\eta}|/\widehat{K}$} & & \\
Data & Method & $0$ & $1$ & $2$ & $\geq3$ &  $0$ & $1$ & $2$ & $\geq3$  & $K-\widehat{K}$ & $d(\widehat{\mathcal{C}},\mathcal{C}^*)$ & comp. time \\ 
			\hline \\[-1.8ex] 
		\multirow{6}{*}{\rotatebox[origin=c]{90}{BLOCKS.out}}  & ebpiece & 0.17 & 0.14 & 0.1 & 0.58 & 0.16 & 0.14 & 0.11 & 0.6 & \textbf{-1.22} &  128.6 & 95.83 (1457.72) \\ 
			 & pelt & 0.1 & 0.52 & 0.11 & 0.27 & 0.05 & 0.25 & 0.05 & 0.65 & -13.8 &  416.6 & \textbf{0.01} \\ 
		& smuce & 0.38 & 0.16 & 0.07 & 0.39 & 0.16 & 0.07 & 0.03 & 0.73 & -15.73 &  301.44 & 0.05 \\ 
			 & solo.cp & 0.4 & 0.15 & 0.06 & 0.39 & \textbf{0.52} & 0.2 & 0.08 & 0.2 & 2.39  & \textbf{108.72} & 112.55 \\ 
			& wbs & \textbf{0.5} & 0.2 & 0.1 & 0.2 & 0.21 & 0.09 & 0.05 & 0.65 & -17.62  & 289.08 & 0.13 \\
			 & r-fpop & \textbf{0.48} & 0.19 & 0.1 & 0.22 & 0.21 & 0.09 & 0.05 & 0.65 & -14.64  & 241.29 & \textbf{0.01} \\ 
			\hline 
			\multirow{6}{*}{\rotatebox[origin=c]{90}{BLOCKS.gauss}}& ebpiece & 0.24 & 0.17 & 0.11 & 0.48 & 0.21 & 0.16 & 0.11 & 0.52 & -1.42 &  100.47 & 99.21 (1484.33) \\ 
			 & pelt & 0.11 & 0.63 & 0.11 & 0.15 & 0.1 & 0.59 & 0.11 & 0.2 & -0.77 & 403.51 & \textbf{0.01} \\ 
			 & smuce & \textbf{0.55}& 0.19 & 0.08 & 0.18 & \textbf{0.6} & 0.21 & 0.08 & 0.1 & 0.93  & 44.12 & 0.05 \\ 
			 & solo.cp & 0.51 & 0.18 & 0.07 & 0.24 & \textbf{0.56} & 0.2 & 0.08 & 0.16 & 0.99  & 81.38 & 113.58 \\ 
			 & wbs & \textbf{0.57} & 0.21 & 0.1 & 0.12 & \textbf{0.57} & 0.21 & 0.1 & 0.12 & \textbf{0} &  17.22 & 0.14 \\ 
 & r-fpop & \textbf{0.59} & 0.21 & 0.09 & 0.1 & \textbf{0.6} & 0.22 & 0.09 & 0.09 & 0.02  & \textbf{12.24} & \textbf{0} \\ 
			 \hline
			\multirow{6}{*}{\rotatebox[origin=c]{90}{BLOCKS.lap}}  & ebpiece & 0.17 & 0.14 & 0.09 & 0.6 & 0.16 & 0.14 & 0.09 & 0.61 & \textbf{-1.24}  & 141.48 & 93.62 (1648.14) \\ 
			 & pelt & 0.1 & 0.46 & 0.13 & 0.3 & 0.09 & 0.41 & 0.12 & 0.39 & -1.85 &  425.79 & \textbf{0.01} \\ 
			& smuce & 0.36 & 0.17 & 0.08 & 0.39 & 0.33 & 0.15 & 0.07 & 0.45 & -1.4 &  179.53 & 0.06 \\ 
		 & solo.cp & 0.34 & 0.14 & 0.06 & 0.47 & \textbf{0.45} & 0.18 & 0.07 & 0.29 & 2.67 &  \textbf{107.91} & 124.1 \\ 
		& wbs & \textbf{0.42} & 0.19 & 0.09 & 0.29 & 0.39 & 0.18 & 0.09 & 0.34 & \textbf{-1.19} & 125.28 & 0.13 \\ 
 & r-fpop & 0.43 & 0.2 & 0.1 & 0.27 & 0.37 & 0.17 & 0.08 & 0.37 & -2.07 & 118.84 & \textbf{0} \\ 
			\hline
			\multirow{6}{*}{\rotatebox[origin=c]{90}{BLOCKS.studt}}  & ebpiece & 0.15 & 0.14 & 0.09 & 0.61 & 0.15 & 0.15 & 0.1 & 0.6 & \textbf{-0.61} & 142.19 & 94.49 (1537.1) \\ 
			 & pelt & 0.1 & 0.48 & 0.13 & 0.29 & 0.07 & 0.33 & 0.09 & 0.51 & -5.92 &  423.5 & 0.01 \\ 
			 & smuce & 0.37 & 0.17 & 0.08 & 0.38 & 0.25 & 0.11 & 0.06 & 0.58 & -6.11 &  255.3 & 0.05 \\ 
		 & solo.cp & 0.35 & 0.15 & 0.06 & 0.44 & \textbf{0.48} & 0.2 & 0.08 & 0.24 & 2.7 &  \textbf{107.07} & 112.32\\ 
			 & wbs & \textbf{0.44} & 0.2 & 0.1 & 0.26 & 0.3 & 0.14 & 0.07 & 0.49 & -6.21 &  215.14 & 0.12 \\ 
 & r-fpop & \textbf{0.45} & 0.21 & 0.1 & 0.25 & 0.29 & 0.14 & 0.07 & 0.51 & -6.71 &  196.87 & \textbf{0} \\ 
			\hline \\[-1.8ex] 
\underline{}	\end{tabular} }
\end{table}

In this section, we consider the \textit{solo.cp} algorithm with $\tau^2_{0,T}=T^{-1}$, $\tau^2_{1,T}=T$, $q=0.1$. For the BLOCKS test signal we set $\tau^2_T=2 T^{-1/2}$ and $\Delta=5$, for the TEETH test signal we set $\tau^2_T=2 T^{-1}$ and $\Delta=2$; the difference is motivated by the smaller sample size of the TEETH data set. For the \textit{basad.cp} algorithm, we employ the default choices of the parameters suggested by \cite{nar14}: $\tau^2_{0,T}= \widehat{\sigma^2} (10 T)^{-1}$, $ \tau^2_{1,T}=\widehat{\sigma^2} \log T$. We use $q=0.1$, and $\Delta=5$ for the BLOCKS signal and $\Delta=2$ for the TEETH signal.
Note that the parameters $\tau^2_{0,T}$, $\tau^2_{1,T}$ are set following the results of Section \ref{sec:theory}. The theory in this paper does not provide guidance on the choice of $\Delta$ and $q$. We study the robustness of the \textit{solo.cp} algorithm to these parameters' choices in Appendix \ref{app:sensi}. 

The procedures \textit{basad.cp}, \textit{solo.cp}, \textit{ebpiece}, and \textit{r-fpop} require the sample standard deviation $\widehat{\sigma}$ as an input. Here, we computed it from the residuals of the fused LASSO \citep{tib05} (implemented in the \texttt{genlasso} \texttt{R} package available on CRAN). The remaining methodologies have an in-built default estimator for $\widehat{\sigma}$. In the BLOCKS signal data sets, we initialize the \textit{ebpiece} MCMC from the estimates of the fused LASSO (``one standard deviation rule" ), otherwise it is not possible to achieve convergence in a reasonable time. This can be seen by the very poor performance of the method which can be due to the fact that the chains ``get stucked" into local modes.
%either to achieve convergence in a reasonable time, or the chains ``get stucked" into local modes.

Tables \ref{tab:block} summarizes $|\widehat{\eta}-\eta|/K$, $|\eta-\widehat{\eta}|/\widehat{K}$, $K-\widehat{K}$, $\widehat{K}$, $d(\widehat{\mathcal{C}},\mathcal{C}^*)$, and the mean computing time for the four scenarios considered for the BLOCKS test signals. Tables \ref{tab:teeth} summarizes the same results for the TEETH test signals. The \textit{basad.cp} method is not included in Table \ref{tab:block} because it was not computationally feasible to approximate the posterior distributions with MCMC in these data sets (the computation time is longer than two hours per data set). %We describe the performance of each method below.  

\begin{table}[!t] \centering 
	\renewcommand{\arraystretch}{0.85} % this reduces the vertical spacing between rows
	\caption{\small{\textbf{Case $n_t=1$:  Haudorff distance, empirical distributions and estimation bias in $K$ of the procedures considered for the TEETH test signals.} Average statistics computed over $100$ simulations for \textit{solo.cp}, \textit{basad.cp}, \textit{ebpiece} \citep{liu20},  \textit{smuce} \citep{fric14},  \textit{wbs} \citep{fryzlewicz2014wild},  \textit{pelt} \citep{killick2012optimal}, and \textit{r-fpop} \citep{fearnhead2018changepoint}. In ``Data", .out refers to mixture of Gaussian errors, .gauss to Gaussian errors, .lap to Laplace errors, and .studt to Student's t errors. For $|\widehat{\eta}-\eta|/K$ and $|\eta-\widehat{\eta}|/\widehat{K}$ the higher the number in the zero column the better. Conversely, for $d(\widehat{\mathcal{C}},\mathcal{C}^*)$ the lower the better. For $K-\widehat{K}$, the closer to the zero the better.  We report in bold the methods with best empirical performance and those within $10\%$ of the best.}}
	\label{tab:teeth} 
	\scalebox{0.8}{
		\hspace{-0.5cm}\begin{tabular}{@{\extracolsep{5pt}} lc|cccc|cccc|c|c|c} 
			\\[-1.8ex]\hline 
			\\[-1.8ex] 
			& &\multicolumn{4}{c}{$|\widehat{\eta}-\eta|/K$} & \multicolumn{4}{c}{$|\eta-\widehat{\eta}|/\widehat{K}$} & & \\
			Data & Method & $0$ & $1$ & $2$ & $\geq3$ &  $0$ & $1$ & $2$ & $\geq3$  & $K-\widehat{K}$  &  $d(\widehat{\mathcal{C}},\mathcal{C}^*)$ & comp. time \\ 
			\hline \\[-1.8ex] 
			\multirow{7}{*}{\rotatebox[origin=c]{90}{TEETH.out}} & basad.cp & 0.78 & 0.1 & 0.04 & 0.08 & \textbf{0.73} & 0.1 & 0.04 & 0.13 & -0.32  & 12.55 & 54 \\ 
			& ebpiece & 0.56 & 0.22 & 0.11 & 0.14 & 0.5 & 0.21 & 0.12 & 0.19 & -0.53 &  \textbf{7.04} & 22.16 \\ 
			& pelt & 0.06 & 0.86 & 0.05 & 0.03 & 0.03 & 0.49 & 0.04 & 0.43 & -3.84 & 21.49 & \textbf{0} \\ 
			& smuce & \textbf{0.8} & 0.11 & 0.06 & 0.03 & 0.5 & 0.08 & 0.05 & 0.37 & -2.91  & 13.65 & 0.04 \\ 
			& solo.cp & 0.76 & 0.06 & 0.04 & 0.13 & \textbf{0.8} & 0.06 & 0.05 & 0.09 & \textbf{0.11}  & 17.09 & 0.05 \\ 
			& wbs & \textbf{0.88} & 0.08 & 0.04 & 0 & 0.42 & 0.06 & 0.04 & 0.48 & -5.72  & 15 & 0.04 \\ 
			& r-fpop & 0.84 & 0.09 & 0.04 & 0.02 & 0.54 & 0.08 & 0.04 & 0.33 & -2.56  & 11.52 & \textbf{0} \\ 
			\hline
			\multirow{7}{*}{\rotatebox[origin=c]{90}{TEETH.gauss}} & basad.cp & \textbf{0.9} & 0.08 & 0.02 & 0 & \textbf{0.87} & 0.08 & 0.02 & 0.04 & -0.2  & 2.91 & 69.49 \\  
			& ebpiece & 0.72 & 0.24 & 0.02 & 0.01 & 0.67 & 0.24 & 0.04 & 0.04 & -0.77  & 3.07 & 23.25 \\ 
			& pelt & 0.02 & 0.96 & 0.02 & 0 & 0.02 & 0.76 & 0.01 & 0.2 & -1.02 & 20.09 & \textbf{0} \\ 
			& smuce & \textbf{0.96} & 0.04 & 0 & 0 & \textbf{0.95} & 0.04 & 0 & 0 & \textbf{-0.02}  & \textbf{0.53} & 0.04 \\ 
			& solo-q0.1 & \textbf{0.94} & 0.04 & 0.01 & 0.02 & \textbf{0.9} & 0.03 & 0.01 & 0.06 & -0.31  & 3.41 & 0.05 \\ 
			& wbs & \textbf{0.96} & 0.04 & 0 & 0 & \textbf{0.93} & 0.04 & 0 & 0.03 & -0.17  & 1.5 & 0.04 \\ 
			& r-fpop & 0.95 & 0.04 & 0.01 & 0 & \textbf{0.87} & 0.04 & 0.01 & 0.08 & -1.33  & 3.34 & \textbf{0} \\ 
			\hline
			\multirow{7}{*}{\rotatebox[origin=c]{90}{TEETH.lap}} & basad.cp & 0.62 & 0.15 & 0.03 & 0.29 & \textbf{0.67} & 0.17 & 0.03 & 0.22 & \textbf{0.37}  & 17.48 & 53.81 \\ 
			& ebpiece & 0.53 & 0.24 & 0.11 & 0.12 & 0.47 & 0.22 & 0.12 & 0.18 & -0.74 & \textbf{6.81} & 22.66 \\ 
			& pelt & 0.08 & 0.78 & 0.07 & 0.07 & 0.06 & 0.6 & 0.05 & 0.29 & -1.27  & 24.39 & \textbf{0} \\ 
			& smuce & \textbf{0.75} & 0.14 & 0.03 & 0.14 & \textbf{0.68} & 0.13 & 0.03 & 0.22 & -0.46  & 7.78 & 0.04 \\ 
			& solo.cp & 0.66 & 0.11 & 0.02 & 0.3 & \textbf{0.74} & 0.13 & 0.03 & 0.19 & 0.51  & 19.44 & 0.05 \\ 
			& wbs & \textbf{0.76} & 0.14 & 0.03 & 0.13 & 0.64 & 0.13 & 0.03 & 0.26 & -0.97 & 8.91 & 0.04 \\
			& r-fpop & \textbf{0.78} & 0.14 & 0.04 & 0.04 & 0.65 & 0.13 & 0.05 & 0.18 & -1.01  & 7.63 & \textbf{0} \\ 
			\hline 
			\multirow{7}{*}{\rotatebox[origin=c]{90}{TEETH.studt}} & basad.cp & 0.72 & 0.12 & 0.04 & 0.13 & \textbf{0.72} & 0.11 & 0.03 & 0.14 & \textbf{-0.08}  & 16.26 & 54 \\ 
			& ebpiece & 0.52 & 0.27 & 0.11 & 0.14 & 0.45 & 0.24 & 0.12 & 0.21 & -0.81 1 & \textbf{6.58} & 22.19 \\ 
			& pelt & 0.05 & 0.82 & 0.1 & 0.03 & 0.03 & 0.54 & 0.06 & 0.37 & -2.68  & 21.04 & \textbf{0} \\ 
			& smuce & \textbf{0.8} & 0.13 & 0.04 & 0.02 & 0.6 & 0.1 & 0.04 & 0.26 & -1.81  & 9.98 & 0.04 \\ 
			& solo.cp & 0.73 & 0.09 & 0.03 & 0.15 &\textbf{0.78} & 0.09 & 0.04 & 0.08 & 0.24 & 16.15 & 0.05 \\ 
			& wbs & \textbf{0.84} & 0.12 & 0.04 & 0.01 & 0.55 & 0.09 & 0.04 & 0.32 & -2.99  & 11.14 & 0.04 \\ 
			& r-fpop & \textbf{0.81} & 0.13 & 0.04 & 0.02 & 0.59 & 0.1 & 0.04 & 0.27 & -1.85 & 10.15 & \textbf{0} \\ 
			\hline \\[-1.8ex] 
	\end{tabular} }
\end{table}

The procedures \textit{wbs}, \textit{smuce}, and \textit{r-fpop} achieve the best overall performance according to $|\eta-\widehat{\eta}|/\widehat{K}$, suggesting that they recover very well $\mathcal{C}^*$; their performance under Gaussian noise scenarios is excellent. The criteria $|\eta-\widehat{\eta}|/\widehat{K}$, $d(\widehat{\mathcal{C}},\mathcal{C})$, and $K-\widehat{K}$ suggest that \textit{wbs} and \textit{smuce} tend to overestimate the number of change points. The problem is extremely severe in the presence of outliers and with Student's t-distributed errors. The method  \textit{r-fpop} is more robust to error specifications and the presence of outliers ($|\eta-\widehat{\eta}|/\widehat{K}$, $d(\widehat{\mathcal{C}},\mathcal{C})$, and $K-\widehat{K}$). However, the biases are still relevant.

\textit{pelt} has generally a good performance but it is worse than \textit{wbs} and \textit {smuce}. It achieves the best performance with Gaussian errors but overestimates $K$. The locations of several change points seem to be shifted by one time instance. However, to the best of our knowledge, the algorithm is implemented correctly. \textit{pelt} and \textit{r-fpop} are the fastest methods employed. 

\textit{ebpiece} recovers well $\widehat{K}$ in both scenarios. It is robust to the misspecification of the error terms ($d(\widehat{\mathcal{C}},\mathcal{C})$ and $K-\widehat{K}$ do not differ much across the four error terms).  It does not seem to recover well the exact locations of the change points ($|\widehat{\eta}-\eta|/K$ and $|\eta-\widehat{\eta}|/\widehat{K}$). In the BLOCKS signals, it is the fastest Bayesian method if the chain is initialized from the output of the fused LASSO. However, the fast computing time in the BLOCKS test signals has to do with the very good initialization employed: if we use the default initialization of the chain used in \citep{liu20}, the procedure takes $30$ min to do the same number of iterations and these are not enough to converge to stationarity (computing time between brackets in Table \ref{tab:block}). It is much slower than \textit{solo.cp} in the TEETH signals. The results are not affected by the initialization in these second data sets. Interestingly, the computing time seems to be mostly affected by the number of MCMC iterations rather than the sample size. %The fast computing time in the BLOCKS test signals has to do with the very good initialization we employ (the output of the fused LASSO): if we employ the random initialization of the chain used in \citep{liu20}, the procedure takes $30$ min to do the same number of iterations and these are not enough to converge to stationarity (computing time between brackets in Table \ref{tab:block}).

\textit{solo.cp} performs well in all scenarios. It is not  as accurate as \textit{wbs} and \textit{smuce} in recovering the exact location of the change points ($|\widehat{\eta}-\eta|/K$). In particular in the BLOCKS scenarios, the reason seems to be that \textit{solo.cp} underestimates $K$. On the other hand, the algorithm is extremely robust to the misspecification of the error terms, being consistently among the best in terms of $|\eta-\widehat{\eta}|/\widehat{K}$ and $d(\widehat{\mathcal{C}},\mathcal{C}^*)$. It is the fastest Bayesian method (accounting for the initialization problem of \textit{ebpiece}). The computing time deteriorates for larger sample size (BLOCKS scenarios). The computing times in the TEETH scenarios are comparable to the state-of-the-art frequentist methods. 

\textit{basad.cp} achieves a performance comparable to \textit{solo.cp} in the TEETH scenarios. This is expected given that both methods are based on the shrinking and diffusing priors of \cite{nar14}. A similar performance is achieved at a much higher computational cost. 

Our overall recommendation is to use either \textit{wbs} or \textit{smuce} in the presence of Gaussian errors and \textit{solo.cp} for non Gaussian errors. An alternative could be to use both procedures and check if there is an agreement in the number of estimated change points. Within the context of Bayesian modeling, \textit{solo.cp} seems to be one of the most viable and accurate procedures available. The results of \textit{solo.cp} are very robust to the choices of $q$ and $\Delta$; see Appendix \ref{app:sensi}. 

\subsection{Case $n_t>1$}
\label{sec:sim_ntlarge}

\begin{figure}[!t]
	\begin{center}
		\includegraphics[scale=0.60]{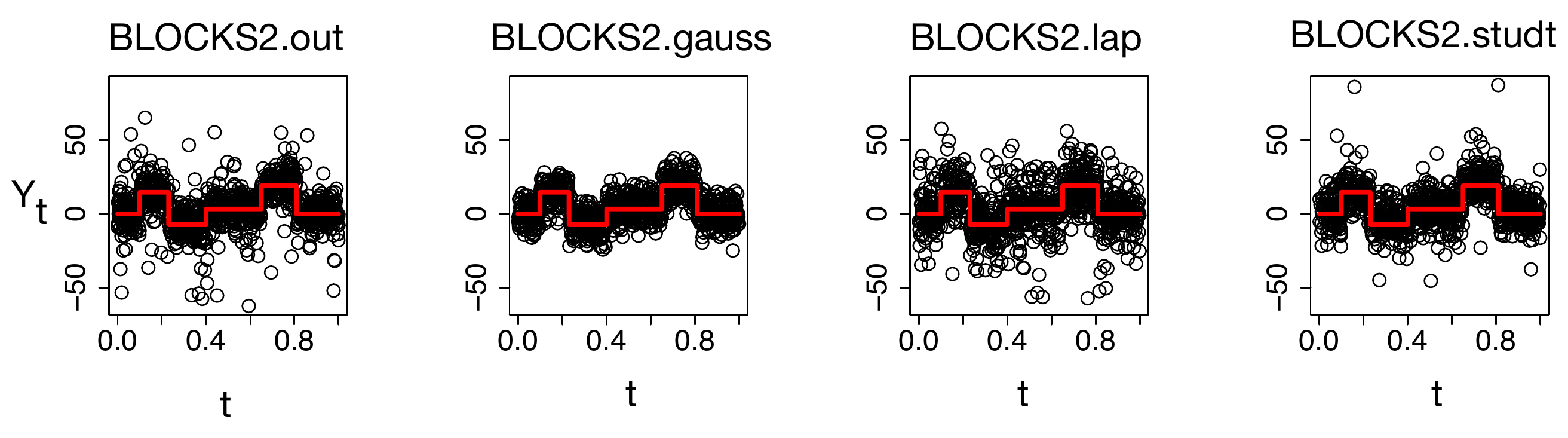}
	\end{center}
	\caption{\small{\textbf{Case $n_t>1$: examples of datasets for the four scenarios considered and true test signal.} First row panels depict the datasets along with the $f_t$ BLOCKS2 (red). First column panels depict observations sampled with mixture of Gaussian errors (.out), the second column panels with the Gaussian errors (.gauss), the third column panels with Laplace errors (.lap), and the fourth column with Student's t errors (.studt).}}
	\label{fig:sim2}
\end{figure}

\begin{table}[!t] \centering 
	\renewcommand{\arraystretch}{0.85} % this reduces the vertical spacing between rows
	\caption{\small{\textbf{Simulations:  Haudorff distance, empirical distributions and estimation bias in $K$ of the procedures considered for the BLOCKS2 test signals.} Average statistics computed over $100$ simulations for \textit{solo.cp}, \textit{basad.cp}, \textit{ebpiece} \citep{liu20},  \textit{smuce} \citep{fric14},  \textit{wbs} \citep{fryzlewicz2014wild}, \textit{pelt} \citep{killick2012optimal}, and \textit{r-fpop} \citep{fearnhead2018changepoint} (the $*$ refers the fact that we are employing local means within each bin).  In ``Data", .out refers to mixture of Gaussian errors, .gauss to Gaussian errors, .lap to Laplace errors, and .studt to Student's t errors. For $|\widehat{\eta}-\eta|/K$ and $|\eta-\widehat{\eta}|/\widehat{K}$ the higher the number in the zero column the better. Conversely, for $d(\widehat{\mathcal{C}},\mathcal{C}^*)$ the lower the better. For $K-\widehat{K}$, the closer to the zero the better.  We report in bold the methods with best empirical performance and those within $10\%$ off the best.}}
	\label{tab:block2} 
	\scalebox{0.8}{
		\hspace{-0.5cm}\begin{tabular}{@{\extracolsep{5pt}} lc|cccc|cccc|c|c|c} 
			\\[-1.8ex]\hline 
			\\[-1.8ex] 
			& &\multicolumn{4}{c}{$|\widehat{\eta}-\eta|/K$} & \multicolumn{4}{c}{$|\eta-\widehat{\eta}|/\widehat{K}$} & & \\\
			Data & Method & $0$ & $1$ & $2$ & $\geq3$ &  $0$ & $1$ & $2$ & $\geq3$  & $K-\widehat{K}$ &  $d(\widehat{\mathcal{C}},\mathcal{C})$ & comp. time \\ 
			\hline \\[-1.8ex] 
			\multirow{7}{*}{\rotatebox[origin=c]{90}{BLOCKS2.out}}  & basad.cp & 0.39 & 0.36 & 0.02 & 0.23 & \textbf{0.49} & 0.45 & 0.02 & 0.05 & 0.94 & 29.91 & 121.86 \\ 
		& ebpiece* & \textbf{0.57} & 0.37 & 0.05 & 0.01 & 0.06 & 0.07 & 0.05 & 0.81 & -42.86 & 33.53 & 73.26 \\ 
			 & pelt* & 0.34 & 0.53 & 0.09 & 0.05 & 0.26 & 0.4 & 0.07 & 0.27 & -1.9 & 39.89 & \textbf{0} \\ 
			 & smuce* & \textbf{0.47} & 0.39 & 0.08 & 0.06 & 0.42 & 0.35 & 0.07 & 0.16 & -0.87 & 8.66 & 0.03 \\ 
			 & solo.cp & 0.44 & 0.44 & 0.03 & 0.09 & \textbf{0.47} & 0.47 & 0.03 & 0.03 & \textbf{0.32} & 15.05 & 0.09 \\ 
			 & wbs* & \textbf{0.48} & 0.4 & 0.08 & 0.04 & 0.4 & 0.34 & 0.07 & 0.19 & -1.57 & \textbf{8.41} & 0.1 \\ 
			 & r-fpop* & \textbf{0.48} & 0.4 & 0.07 & 0.05 & 0.34 & 0.3 & 0.07 & 0.29 & -2.81 & 13.35 & \textbf{0} \\ 
			\hline
			\multirow{7}{*}{\rotatebox[origin=c]{90}{BLOCKS2.gauss}} & basad.cp & 0.53 & 0.44 & 0.01 & 0.02 & \textbf{0.53} & 0.45 & 0.01 & 0.01 & 0.05 & 4.79 & 122.7 \\ 
		 & ebpiece* & \textbf{0.6} & 0.37 & 0.02 & 0.01 & 0.07 & 0.07 & 0.05 & 0.81 & -39.58 & 31.62 & 67.14 \\ 
		 & pelt* & 0.36 & 0.55 & 0.07 & 0.02 & 0.24 & 0.38 & 0.06 & 0.33 & -2.88 & 39.48 & \textbf{0} \\ 
			 & smuce* & 0.51 & 0.41 & 0.04 & 0.04 & 0.38 & 0.31 & 0.04 & 0.27 & -1.94 & 11.06 & 0.04 \\ 
			 & solo.cp & \textbf{0.54} & 0.44 & 0.01 & 0 & \textbf{0.54} & 0.44 & 0.01 & 0 & \textbf{0} & \textbf{2.2} & 0.08 \\ 
			 & wbs* & 0.53 & 0.4 & 0.04 & 0.03 & 0.35 & 0.29 & 0.05 & 0.3 & -3.22 & 11.24 & 0.1 \\ 
			 & r-fpop* & 0.53 & 0.41 & 0.04 & 0.02 & 0.29 & 0.24 & 0.05 & 0.41 & -5.14 & 16.05 & \textbf{0} \\ 
			\hline
		\multirow{7}{*}{\rotatebox[origin=c]{90}{BLOCKS2.lap}} & basad.cp & 0.32 & 0.26 & 0.01 & 0.45 & 0.54 & 0.41 & 0.02 & 0.07 & 1.95 & 36.54 & 120.36 \\ 
 & ebpiece* & \textbf{0.53} & 0.38 & 0.06 & 0.03 & 0.08 & 0.09 & 0.06 & 0.77 & -27.17 & 29.82 & 59.51 \\ 
& pelt* & 0.33 & 0.52 & 0.09 & 0.05 & 0.26 & 0.42 & 0.08 & 0.24 & -1.49 & 40.07 & \textbf{0} \\ 
 & smuce* & 0.46 & 0.38 & 0.08 & 0.08 & \textbf{0.43} & 0.36 & 0.08 & 0.13 & \textbf{-0.43} & \textbf{7.65} & 0.04 \\ 
 & solo.cp & 0.41 & 0.38 & 0.05 & 0.16 & \textbf{0.46} & 0.44 & 0.05 & 0.04 & 0.57 & 23.55 & 0.1 \\ 
 & wbs* & \textbf{0.48} & 0.39 & 0.07 & 0.06 & 0.4 & 0.35 & 0.07 & 0.17 & -1.32 & \textbf{8} & 0.11 \\ 
 & r-fpop* & 0.47 & 0.38 & 0.08 & 0.07 & 0.38 & 0.31 & 0.08 & 0.22 & -1.7 & 11.24 & \textbf{0} \\ 
\hline
			\multirow{7}{*}{\rotatebox[origin=c]{90}{BLOCKS2.studt}} & basad.cp& 0.45 & 0.36 & 0.02 & 0.16 & \textbf{0.52} & 0.42 & 0.03 & 0.03 & 0.68 & 24.01 & 124.82 \\ 
		 & ebpiece* & \textbf{0.6} & 0.36 & 0.03 & 0.01 & 0.07 & 0.07 & 0.05 & 0.81 & -40.82 & 33.22 & 65.37 \\ 
		 & pelt* & 0.35 & 0.55 & 0.06 & 0.04 & 0.26 & 0.41 & 0.05 & 0.28 & -2.02 & 39.71 & \textbf{0} \\ 
			 & smuce* & 0.49 & 0.39 & 0.07 & 0.05 & 0.42 & 0.34 & 0.06 & 0.17 & -0.96 & \textbf{8.67} & 0.04 \\ 
			 & solo.cp & 0.49 & 0.43 & 0.02 & 0.05 & \textbf{0.51} & 0.45 & 0.02 & 0.02 & \textbf{0.16} & 10.26 & 0.08 \\ 
		 & wbs* & 0.51 & 0.39 & 0.06 & 0.04 & 0.4 & 0.31 & 0.06 & 0.24 & -2.1 & 9.79 & 0.09 \\
		& r-fpop* & 0.5 & 0.39 & 0.06 & 0.05 & 0.34 & 0.29 & 0.07 & 0.3 & -2.93 & 13.03 & \textbf{0} \\ 
			\hline \\[-1.8ex] 
			\underline{}	\end{tabular} }
\end{table}

We now consider situations where the number of data points collected at any time instance can be more than one. %This study mimics two realistic scenarios: as part of the sampling design, one collects multiple observations at a given time point; the analyst is interested in inferring the time intervals when change points happened. In this second case, observations could be not equally spaced. 
We use a test signal called BLOCKS2 ($K=6, n=1024$), which is a simplified version of BLOCKS. We cannot employ BLOCKS because binning observations into an equally spaced grid results in change points that are too close to each other. We employ the four error distributions used in Section \ref{sec:sim_nt1}. Details are give in Appendix \ref{app:sim.det}. To generate each dataset we sample $n$ time points uniformly at random on the interval $[0,1]$. Then we sample each observation at the corresponding $t$ from $f_t$ (\textit{i.e.} the observations are not equally spaced). Finally, we bin all the observations according to a regular grid on $[0,1]$ with $200$ intervals. Figure \ref{fig:sim2} depicts examples of possible datasets along with the BLOCKS2 signal (red line).

Our methods naturally allow for this setting. The parameters of \textit{basad.cp} and \textit{solo.cp} are set equal to the ones used for the BLOCKS signal in Section \ref{sec:sim_nt1}. Note that in this case, we use the number of grid points instead of $T$ to define the parameters. The other methods are not designed for this setting. We compute the local means within each bin and feed the local means to each method. The rest of the simulation setup is identical to Section \ref{sec:sim_nt1}. Table \ref{tab:block2} summarizes the results. We describe the performance of each method below. 

The methods \textit{r-fpop}, \textit{wbs}, \textit{smuce} and \textit{pelt} perform well in all the scenarios. However, they are no longer the best performing methods in terms of $|\widehat{\eta}-\eta|/K$. They do not seem as sensitive, as in the case $n_t=1$, to the misspecification of the error terms. We hypothesize that this follows from the use of the local means which make outliers less relevant. This is signaled by the low value of Hausdorff metric. However, the tendency to overestimate $K$ remains prevalent, as suggested by the statistic $K-\widehat{K}$.

The performance of \textit{ebpiece} is better than in the previous section. It is often the best method in terms of $|\widehat{\eta}-\eta|/K$, which indicates that the locations of the change points are correctly recovered. This happens because the number of change points is severely overestimated ($K-\widehat{K}$). 

Our procedure \textit{solo.cp} has the best overall performance in several metrics across scenarios. The method remains robust to misspecified error terms. Furthermore, the algorithm is extremely competitive also under Gaussian errors. The computing time is in line with the alternatives. \textit{basad.cp} achieves a very similar performance but with a much higher computational cost.

Overall, our recommendation is to use \textit{solo.cp} for univariate mean change point detection. 

\section{Applications}
\label{sec:app}

\subsection{Array Comparative Genomic Hybridization (aCGH) data}

Genomic alternations happen in the development of tumors. Studying these alternations, for example determining the copy-number variations, is important for understanding cancer and also used for its diagnosis. Array Comparative Genomic Hybridization (aCGH) is a popular method that generates this type of data \citep{sche95}. We analyze an aCGH dataset of individuals with a bladder tumor collected by \cite{stra06}. The dataset is publicly available in the \texttt{R} package \texttt{ecp} \citep{jam14}, and includes $43$ individuals and $2215$ locations. The goal of the analysis is to detect changes in the copy-number. The underlying assumption is that alternations are constant within a segment. Segments involved in the tumor should be equally affected across patients. 

\begin{figure}[!t]
	\begin{center}
		\includegraphics[scale=0.58]{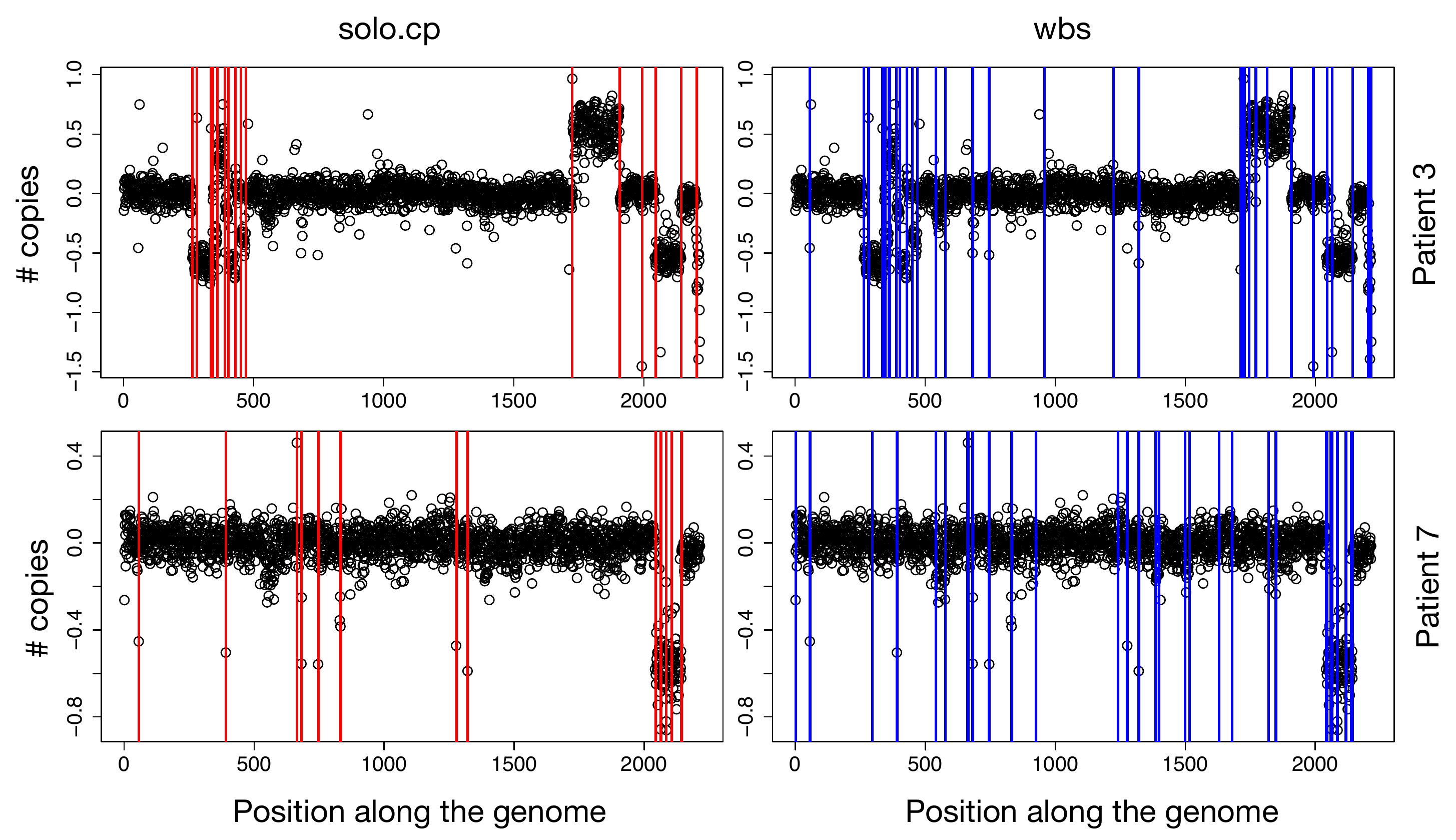}
	\end{center}
	\caption{\small{\textbf{aCGH data.} Copy number variations of patients having a bladder tumor recorded at $n=2215$ sites by \cite{stra06}. First row depicts the observations for Patient $3$, second row depicts the observations for Patient $7$. First column includes the change points estimated by \textit{solo.cp} (red), second column the ones estimated by \textit{wbs} (blue).}}
	\label{fig:acgh}
\end{figure}

While we could repeat the analysis for all the patients, we include only two in this manuscript for parsimony. The number of samples is approximately identical to the BLOCKS test signal, hence we use the same parameters for \textit{solo.cp} ($\tau^2_{0,T}=T^{-1}$, $\tau^2_{1,T}=T$, $\tau^2_{T}=T^{-1/2}$, $q=0.1$, $\Delta=5$) and $\widehat{\sigma^2}$ equal to the variance of the residuals of the fused LASSO (tuning parameter chosen by one-standard-error rule). We compare the results of \textit{solo.cp} with \textit{wbs} (default implementation). The left column of Figure \ref{fig:acgh} depicts the estimates of the \textit{solo.cp} change points, the right column depicts the one obtained with \textit{wbs}. The two rows refer to the two different patients.

Both methods seem to recover more change points than the number of blocks identified through a visual inspection of the data. There are a few points where the change point corresponds to a single observation, not an entire segment along the genome. We would need further research to determine if these points can be classified as outliers. However, we note that \textit{solo.cp} appears more parsimonious:  $\widehat{K}=19$ for Patient $3$ and $\widehat{K}=13$  for Patient 7, while \textit{wbs} estimates $\widehat{K}=49$ for Patient $3$ and $\widehat{K}=36$  for Patient $7$. The results are consistent with what we observed in the simulation section.

\subsection{Ion channels data}

Ion channels are a class of proteins expressed by all cells that create pathways for ions (charged particles) to pass through the otherwise impermeable cell membrane. The opening of these pathways is essential for cell operations in the nervous system, in the muscles, and in the pancreas. Thus, the study of ion channels plays a fundamental role in the development of new drugs \citep{ale08}. The patch clamp technique is an electrophysiological tool for understanding ion channel behavior. It is used to measure ionic currents from single living cells or tissues \citep{neh95}. Electrophysiologists use glass microelectrodes to gain access to cells expressing ion channels. Through the microelectrode, a voltage is applied, forming a voltage clamp, and the current passing across the cell membrane through the ion channels is measured. 
%The path clamp technique allows researchers to measure ion currents through the membrane of a single cell \citep{neh95}. The technique involves ``blocking" the voltage potential in an area of a cell membrane in order to measure the variation in voltage induced by the ion channels in that area. 

\begin{figure}[!t]
	\begin{center}
		\includegraphics[scale=0.60]{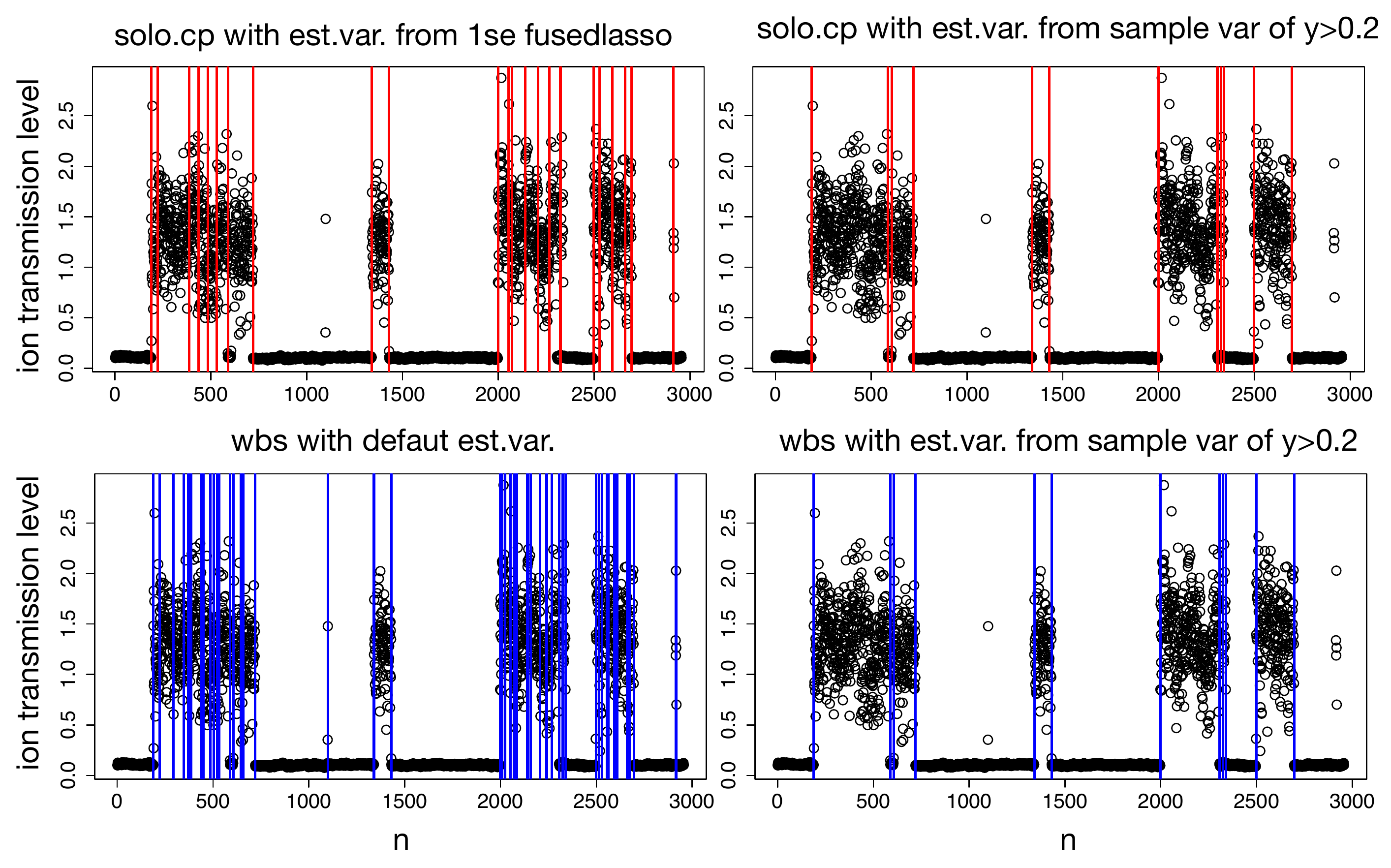}
	\end{center}
	\caption{\small{\textbf{Ion data.} Ion channel data recorded at
the Steinem lab (Insitute of Organic and Biomolecular Chemistry, University of G{\"o}ttingen) at $n=2956$ time instances. First row depicts the change points estimated through \textit{solo.cp}, second row depicts  the change points estimated through \textit{wbs}. Estimates in the first column are obtained using ``default" estimates of the sample standard deviation, which means the standard deviation of the residuals obtained from the fused LASSO (one standard error rule) for \textit{solo.cp}, and the median absolute deviation estimates for \textit{wbs}. Estimates in the second column are obtained using the sample standard deviation of observations taking values larger than $0.2$ (roughly speaking to approximate when ion channels are open). }}
	\label{fig:ion}
\end{figure}

We consider a dataset produced by the Steinem Lab (Institute of Organic and Biomolecular Chemistry, University of G{\"o}ttingen), recently analyzed by \cite{vanegas2019multiscale}, measuring a single ion channel of the bacterial porin PorB, a bacterium that plays a role in the pathogenicity of Neisseria gonorrhoeae. The experiment design includes a technique that induces local dependencies of the error terms \citep{pei17}. To remove these dependencies, we follow the same approach of \cite{vanegas2019multiscale}, subsampling every $11$th observation. The original dataset includes $600000$ time instances. We analyze a portion of the dataset of length $32511$. After subsampling, the data set is composed of $2956$ time points. Figure \ref{fig:ion} depicts the data set. %The same analysis can be repeated for every segment. 

Figure \ref{fig:ion} suggests that the noise variance  when the channels are open is much higher than when they are closed. This feature of ion channel data is known as \textit{open channel noise} \citep{neh95}. The methods studied in this paper, and considered in Section \ref{sec:sim}, do not assume  error heterogeneity. The first row of Figure \ref{fig:ion} depicts the estimated change points of \textit{solo.cp} ($\tau^2_{0,T}=T^{-1}$, $\tau^2_{1,T}=T$, $\tau^2_{T}=T^{-1/2}$, $q=0.1$,$\Delta=5$) and $\widehat{\sigma^2}=.037$ being the variance of the residuals of the fused LASSO (tuning parameter chosen by one-standard-error rule), and \textit{wbs} run with its default setting. \textit{solo.cp} estimates $23$ change points, \textit{wbs} estimates $48$. A visual inspections of the plot suggests that some of the estimates might be redundant. The second row depicts the estimates obtained with \textit{solo.cp} using the same parameters and $\widehat{\sigma^2}=.137$ being the sample variance of the observations when the ion channels are open (we approximate it considering observations above $0.2$). \textit{wbs} is also run with this $\widehat{\sigma^2}$ (it is also require an extra parameter $\theta_0$, here chosen equal to $3$). Now, $\widehat{K}$ is $12$ for both methods and the locations of the change points seem reasonable by visual inspection. A few isolated points are not detected as change points (approximately around $1100$ and $2900$). We note though that the result of \textit{wbs} largely depends on other tuning parameters (\textit{e.g} $\theta_0$), while the estimates of \textit{solo.cp} are very robust to the choices of all the parameters that are not  $\widehat{\sigma^2}$.

\section{Discussion}
\label{sec:disc}

We studied spike and slab priors for change point detection leveraging recent results in the variable selection literature. We chose to work with a prior having both the spike and the slab component defined by Gaussian distributions and sample size-dependent hyperparameters.  We established that an estimator based on this prior distribution is consistent and achieves optimal localization rates of multiple change points. Furthermore, the use of this prior allowed us to propose a fast Bayesian change point estimator based on a slightly misspecified model. A version of the fast estimator achieves the optimal rate in the single change point problem. In simulations, its empirical accuracy is comparable to state-of-the-art benchmarks. Its salient features are being one of the fastest Bayesian methods available (no MCMC required) and being very robust to misspecification of the error model. We showed these features in simulation studies, displaying situations where our estimator performs well while many competing methods severely overestimate the number of change points. 

There is a rich literature on change point detection for settings more general than the one considered in this paper. Nevertheless, our results are promising and suggest that it is worth investigating the use of spike and slab priors in change point detection for more general settings, such as settings with unknown variance, heterogeneous errors, and different types of dependence. 

The first area of future work is to further improve the computational performance of the \textit{solo.cp} algorithm. The main bottleneck of the algorithm is the computation of the parameters in \eqref{eq:post.param}. The computing time of \textit{solo.cp} is comparable to those of frequentist estimators for small sample sizes (approx $n=200$), but it deteriorates for large sample sizes (in the order of the thousands). % Computing these parameters explain the differences

The second area of research is the detection of higher-order changes, such as in piecewise-linear signals. A version of the \textit{solo.cp} algorithm for piecewise-linear change point detection is readily available (as well as higher-order changes). However, our preliminary results suggest that a vanilla version of this estimator does not work well in this setting. \cite{liu20} suggest that a possible explanation is that one cannot fix the prior means at zero in this setting.

\appendix

\section{Extension of  \textit{solo.cp}  to the case $n_t>1$}
\label{app:solo.cpnt>1}

Let $(y_{1,t}, \ldots, y_{n_t,t})$ denote the vector of $n_t$ observations collected at time $t$, with $t$ in $1$ to $M$. $T$ is the total sample size $T=\sum_{t=1}^M n_t$. The extension of \textit{solo.cp} to this setting is straightforward: the spike and slab priors will be placed on the $M$ piecewise changed $\Delta f_t$, and parameters \eqref{eq:singn_para1}, \eqref{eq:singn_para2}, and \eqref{eq:post.param} need to be adjusted to account for the fact that multiple observations are collected at a given time point. 

Below, we provide the equivalent of parameters \eqref{eq:singn_para1}, \eqref{eq:singn_para2}, and\eqref{eq:post.param}. Suppose we are interested in testing whether $\Delta f_j$ is a change point. Initialize $n'_{M}=(\tau^2_n n^2_M)/(\tau_n^2 n_M + \sigma^2)$, $\overline{y'_{M}}=\sum_{i=1}^{n_M} y_{i,M}/n_M$, then for $i=M-1$ to $j+1$ compute
\begin{align}
\label{eq:mutn_para1}
n'_{i}&=\frac{\tau_M^2 (\sum_{k=i}^M n_k - \sum_{k=i+1}^M n'_{k})^2}{[\tau_M^2 \sum_{k=i}^M n_k - \sum_{k=i+1}^M n'_{k}+\sigma^2]} \,\ \,\,\, \text{and} \,\ \,\,\, \overline{y'_{i}}=\frac{\sum_{t=i}^M \sum_{k=1}^{n_t} y_{k,t} - \sum_{k=i+1}^M n'_{k} \overline{y'_{k}}}{\sum_{k=i}^M n_k - \sum_{k=i+1}^M n'_{k}}.
\end{align}
Then, set $\gamma_{j,1}=1$, and for $i=1$ to $j$ compute
\begin{align}
\label{eq:mutn_para2}
n''_{i,j} &=\frac{\tau_M^2}{\tau_M^2 \gamma_{i,j} (\sum_{k=i}^M n_k - \sum_{k=j+1}^M n'_{k})+\sigma^2}, \nonumber\\
\overline{y''_{i,j}}&=\sum_{t=i}^M\sum_{k=1}^{n_t} y_{k,t} - \sum_{k=j+1}^M n'_{k} \overline{y'_{k}}- \Big(\sum_{k=i}^M n_k- \sum_{k=j+1}^Mn'_{k}\Big)\Big[\sum_{k=1}^{i-1} n''_{k,j} \gamma_{k,j} \overline{y''_{k,j}} \Big],\\
\gamma_{i+1,j}&=1-\Big(\sum_{k=i}^M n_k- \sum_{k=j+1}^Mn'_{k}\Big)\Big(\sum_{k=1}^{i} n''_{k,j} \gamma_{k,j}^2 \Big). \nonumber	
\end{align}
Finally, the posterior parameters, for $k \in \{0,1\}$:
\begin{align}
\label{eq:multn_postparam}
\mu_{k,j}&=\frac{\overline{y''_{j,j}}}{(\sum_{k=j}^Mn_k - \sum_{k=j+1}^Mn'_{k})\gamma_{j,j} +\tau_{k,M}^{-2} \sigma^2},\\
\xi_{k,j}&=\frac{\sigma^2}{(\sum_{k=j}^M n_k - \sum_{k=j+1}^Mn'_{k})\gamma_{j,j} +\tau_{k,M}^{-2} \sigma^2}, \nonumber\\
\omega_{0,j} &=\exp \Bigg\{\frac{1}{2 \sigma^2} \frac{\overline{y_{j,j}^{''}}^2}{(\sum_{k=j}^M n_k - \sum_{k=j+1}^Mn'_{k})\gamma_{j,j} +\tau_{k,M}^{-2} \sigma^2} \Bigg\}
 \times \sqrt{\frac{\tau_{k,M}^{-2}}{(\sum_{k=j}^M n_k - \sum_{k=j+1}^Mn'_{k})\gamma_{j,j} +\tau_{k,M}^{-2} \sigma^2}}.\nonumber
\end{align}
The rest of the procedure continues as described in Algorithm \ref{all:all}.

\section{Notation for proofs}

For two  sequences $a_n$  and $b_n$  we write $a_n \lesssim b_n$ if there exits a constant  $C>0$ such that $a_n \leq  C b_n$  for all $n$. Similarly, we denote $a_n \gtrsim b_n$  if   there exits a constant  $C>0$ such that $a_n \geq  C b_n$  for all $n$.

The sampling model can be rewritten as
\begin{equation}
	\label{eqn1=2}
	\textbf{Y} \,=\, X(T) \, \bm{\Delta f} \,+\, \epsilon,
\end{equation}
where  $\epsilon \sim N(0,\sigma^2 I_T)$, $I_T \in \mathbb{R}^{T\times T}$ is the identity matrix and 
\begin{equation}
	\label{eqn:matrix}
	X(T)   \,=\,\left(\begin{matrix}  
		1 & 0  &0 &0    &\ldots  &  0 & 0\\
		1 & 1 & 0  & 0  &\ldots & 0 & 0\\
		1 & 1& 1  & 0  & \ldots  & 0 &0 \\
		\vdots&	\vdots&  	\vdots&  	\vdots&  \ldots  &\vdots &\vdots\\
		1 & 1&1&1 & \ldots & 1 & 1  
	\end{matrix}\right) \in \mathbb{R}^{T \times T}.
\end{equation}
%$\beta_1^* =  f_1$, and  $\beta_j^*  =  f_{j} - f_{j-1} $  for  all $j \in \{2,\ldots,T\}$.
%$\mathcal{C}^* = \{ j\,\,:\,   \beta_j^* \neq 0 \}$ .  Then $\mathcal{C}^*\backslash \{\beta^*_1\}$  is the set of jumps in the mean of the sequence vector $\textbf{Y} \in \mathbb{R}^T$. 

With the notation from (\ref{sec:th1})  we can write  (\ref{eqn:bayesian}) as 
\begin{equation}
	\label{eqn:bayesian2}
	\begin{array}{lll}
		\widetilde{\textbf{Y}} \,|\,    X(m) , \bm{\Delta f},\sigma^2  & \sim &     \mathcal{N}\left(    X(m)  \bm{\Delta f}, \sigma^2 I_{m}  \right),\\
		\Delta f_j\,|\,   \sigma^2, Z_{j} = 0 ,\tau_{0,m}^2   &\sim &     \mathcal{N}\left(   0, \sigma^2   \tau_{0,m}^2 \right),\\
		\Delta f_j   \,|\, \sigma^2, Z_{j} = 1,\tau_{1,m}^2    &\sim &     \mathcal{N}\left(   0, \sigma^2   \tau_{1,m}^2 \right),\\
		P(Z_j= 1)    &\, =\,&   1- P(Z_j = 0) =  q_m,\,\,\,\,\,\,\, j =1 \,\ldots,,m,\\
		%	\sigma^2   & \sim&   IG(\alpha_1,\alpha_2),
	\end{array}      
\end{equation}
where  $X(m) \in \mathbb{R}^{m \times m}$  is the corresponding matrix  in (\ref{eqn:matrix})  but with   $m$ instead of $T$, and where  $q_m ,\tau_{0,m},\tau_{1,m}>0$.
% Notice that $\beta_j$ represents a parameter for the jump size at location $j$.% \gr{It is equivalent to the piecewise constant change $\Delta f_j$ used in the previous section. }% Thus for all 

For the proof of Theorem \ref{thm2} we use the following notation. As  in \cite{che19}, we denote by  $x_j$ the $j$th column of the matrix $X = X(T)$, and $X_{-j}$ the matrix obtained after removing the $j$th column of  $X$.  We then consider the Bayesian model (\ref{eq:solo.cp}) for a fixed  $j \in \{1,\ldots,T\} $, which can be written as
\[
\begin{array}{lll}
	\textbf{Y}  \,|\,     X,\Delta f_j, \bm{\Delta f_{-j}},\sigma^2 &  \sim&   N(   \Delta f_j x_j  +      X_{-j} \bm{\Delta f_{-j}} ,\sigma^2 I_T)\\
	\Delta f_j\,|\,   \sigma^2, Z_{j} = 0 ,\tau_{0,m}^2   &\sim &     \mathcal{N}\left(   0, \sigma^2   \tau_{0,m}^2 \right),\\
	\Delta f_j   \,|\, \sigma^2, Z_{j} = 1,\tau_{1,m}^2    &\sim &     \mathcal{N}\left(   0, \sigma^2   \tau_{1,m}^2 \right),\\
	\Delta f_{i} | \sigma^2, \tau^2_{T} &\sim& \mathcal{N}(0, \sigma^2 \tau^2_{T}),     \,\,\,\,\,\,\, i\in \{1 \,\ldots,T\}\backslash \{j\},    \\
	P(Z_j=1)&=& 1- P(Z_j=0)\,=\, q_{j,T},\\
\end{array}
\]
%Again, $\beta_j$ corresponds to the piecewise increment $\Delta f_j$. 
Then from Section 3  from \cite{che19} we obtain that 
%    and notice that as in \cite{che19}, it holds that 
\[
%\mathbb{E}\left(    \right)
P(  \Delta f_j   \, |\,\sigma^2, \textbf{Y})      \,\propto \, (1-q_{j,T}) w_{0,j} N(\Delta f_j  \,|\,    \mu_{0,j}, \varepsilon_{0,j}^2)   \,+\, q_{j,T}w_{1,j}  N(\Delta f_j  \,|\,    \mu_{1,j}, \varepsilon_{1,j}^2)  
\]
where for  $k\in \{0,1\}$ we have that 
\[
\mu_{k,j}\,:=\, \frac{x_j^{\top }( I - \tilde{H}_j)\textbf{Y}  }{ x_j^{\top }( I - \tilde{H}_j)x_j +   \tau_{k,T}^{-2}      },
\]
\[
\varepsilon_{k,j}^2 \,:=\,   \frac{\sigma^2}{   x_{j}^{\top}(I- \tilde{H}_j)x_j + \tau_{k,T}^{-2}    }
\]
with 
\[
\tilde{H}_j  \,:=\,   X_{-j}\left( X_{-j}^{\top} X_{-j}+   \tau_{T}^{-2}I    \right)^{-1}X_{-j}^{\top},
\]
and for some  positive weights   $w_{0,j}$ and  $w_{1,j}$.

Furthermore, as  \cite{che19} showed, the posterior means $\{\mu_{1,j}\}$ satisfy 
\[
\mu_{1,j}   \,\sim\,   N\left(        \frac{  x_j^{\top }( I - \tilde{H}_j)X \bm{\Delta f}}{   x_j^{\top }( I - \tilde{H}_j)x_j +   \tau_{1,T}^{-2}   }  ,    \sigma^2\frac{ x_j^{\top }( I - \tilde{H}_j)^2x_j}{(x_j^{\top }( I - \tilde{H}_j)x_j +   \tau_{1,T}^{-2})^2} \right).
\]

\section{Auxiliary lemmas for  proof of Theorem \ref{thm1} }

\begin{lemma}
	\label{lem1}
	Let $X  \in \mathbb{R}^{T \times T}$ be the matrix defined in (\ref{eqn:matrix}), $\lambda_{\min}(\cdot) $ denote the minimum eigenvalue function, and $\lambda_{\max}(\cdot) $ denote the maximum eigenvalue function.  Then
	\[
	\lambda_{\min} \left(   \frac{X^{\top}X}{T} \right)\,\geq \,  \frac{1}{4T}.
	\]
\end{lemma}
\begin{proof}
Notice that 
\[
\lambda_{\min}\left(  \frac{X^{\top} X }{  T }  \right) \,=\,    \frac{  \lambda_{\min}\left(   X^{\top}X  \right)   }{T}\,=\,   \frac{ 1}{T   \lambda_{\max}\left(   \left(X^{\top} X\right)^{-1}  \right)  }.
\]	
Furthermore,  as in the supplementary material of \cite{tibshirani2014adaptive}, one can verify that 
\[
\left(X^{\top} X\right)^{-1} \,=\, \left(  \begin{matrix}
	2 & -1  & 0 & 0&0 &\ldots  &0  \,\,\,\,\,\,0 &0&0 &0 \\%&0&0&0&0\\
	-1& 2 &-1 & 0 &0 &\ldots   &0  \,\,\,\,\,\,0 &0&0 &0 \\
	0 &-1&2&-1&0 & \ldots    &0  \,\,\,\,\,\,0 &0&0 &0 \\
	\vdots&\vdots&\vdots&\vdots&\vdots&\ldots& \vdots  \,\,\,\,\,\ \vdots& \vdots& \vdots& \vdots\\
	0 &0&0&0&0 & \ldots    &0  \,\,\,\,\,\,-1 &2&-1 &0 \\
	0 &0&0&0&0 & \ldots    &0\,\,\,\,\,\,0&-1& 2&-1 \\	
	0 &0&0&0&0 & \ldots    &0\,\,\,\,\,\,0&0& -1&1 \\	
	%   	0&0&0&   	
\end{matrix}  \right) \,\in \mathbb{R}^{T \times T}.
\] 
However, 
\[
\lambda_{\max}\left(\left(X^{\top} X\right)^{-1} \right)  \leq      \lambda_{\max}\left(   A  \right)   \,+\, \lambda_{\max}(B),
\]
where  
\[
A\,=\, \left(  \begin{matrix}
	2 & -1  & 0 & 0&0 &\ldots  &0  \,\,\,\,\,\,0 &0&0 &0 \\%&0&0&0&0\\
	-1& 2 &-1 & 0 &0 &\ldots   &0  \,\,\,\,\,\,0 &0&0 &0 \\
	0 &-1&2&-1&0 & \ldots    &0  \,\,\,\,\,\,0 &0&0 &0 \\
	\vdots&\vdots&\vdots&\vdots&\vdots&\ldots& \vdots  \,\,\,\,\,\ \vdots& \vdots& \vdots& \vdots\\
	0 &0&0&0&0 & \ldots    &0  \,\,\,\,\,\,-1 &2&-1 &0 \\
	0 &0&0&0&0 & \ldots    &0\,\,\,\,\,\,0&-1& 2&-1 \\	
	0 &0&0&0&0 & \ldots    &0\,\,\,\,\,\,0&0& -1&2 \\	
	%   	0&0&0&   	
\end{matrix}  \right) \,\in \mathbb{R}^{T \times T}
\] 
and 
\[
B  \,=\,  \left(   \begin{matrix}0& \ldots& 0&0\\ \vdots & \ddots & \vdots&  \vdots \\ \vdots& &0&0\\0 &\ldots&0&-1\end{matrix}\right)  \in \mathbb{R}^{T \times T}.
\]
Therefore, 
\[
\lambda_{\max}\left(\left(X^{\top} X\right)^{-1} \right)  \leq    \lambda_{\max}(A) \,\leq \, 4,
\]
since the eigenvalues of  $A$ are 
\[
2 +2 \mathrm{cos}\left( \frac{k\pi}{T+1} \right),\,\,\,\, k =1,\ldots,T,
\]
which holds by the fact that $A$ is a tridiagonal Toeplitz matrix.
\end{proof}

\begin{lemma}
	\label{lem3}
	Let $X  \in \mathbb{R}^{T \times T}$  the matrix defined in (\ref{eqn:matrix}).  Then
	\[
	\lambda_{\max} \left(   \frac{X^{\top}X}{T} \right)\,\leq \,  2T.
	\]
\end{lemma}
\begin{proof}
	Let  $v \in \mathbb{R}^T$  with  $\|v\|=1$. Then
	\[
	\begin{array}{lll}
		\displaystyle v^{\top}X^{\top} Xv&  = &  \displaystyle  v_1^2  +    ( v_1+v_2)^2   + \ldots + (v_1 +v_2+\ldots+v_T)^2  \\
		& = &\displaystyle v_1^2   + (v_1^2 +   2v_1 v_2 +v_2^2   ) +\ldots + \left(v_1^2 +       \ldots  +v_T^2 +      2 \sum_{i\neq j} v_i v_j  \right) \\
		& \leq&   \displaystyle   v_1^2   + (v_1^2 +   (v_1^2+ v_2^2) +v_2^2   ) +\ldots + \left(v_1^2 +       \ldots  +v_T^2 +       \sum_{i\neq j} (v_i^2+ v_j^2)  \right)\\
		& \leq& 2T^2 \left(   v_1^2 +       \ldots  +v_T^2\right),
	\end{array}  
	\]
	where the first inequality follows from Cauchy-Schwarz inequality, or $a^2+b^2-2ab\geq0$.
\end{proof}

Recall that  $\Lambda_1,\ldots,\Lambda_m$  is a partition of  $\{1,\ldots,T\}$ such that  $\vert \Lambda_j \vert    \asymp    T/m$ for all $j \in \{1,\ldots,m\}$, with $m \in \mathbb{N}$  with  $m \leq  T$.
\begin{lemma}
	\label{lem2}
	Let   $\tilde{f}_j \,=\,  \mathbb{E}( \tilde{Y}_j)$  for  $j \in \{1,\ldots,m\}$  and assume $m \,\asymp\, \frac{\kappa^2   T }{\sigma^2\log T}$, then for  $j\in\{2,\ldots,m\}$ the following holds:
	\begin{itemize}
		\item If  $\mathcal{C}^*  \cap \Lambda_{j-1} =\emptyset$  and $\mathcal{C}^*  \cap \Lambda_j =\emptyset$   then $\tilde{f}_j = \tilde{f}_{j-1}$.
		\item If  $  \mathcal{C}^*  \cap \Lambda_{j}  \neq \emptyset$     then 
		\[
		\max\{\vert \tilde{f}_j - \tilde{f}_{j-1}  \vert,\vert \tilde{f}_{j+1}- \tilde{f}_{j}  \vert \} \,\geq \, \sqrt{\frac{T}{m}} \frac{\kappa}{2}.
		\]
%		for a positive constant $c_0>0$.
	\end{itemize}
\end{lemma}

\begin{proof}
	First notice that  by Assumption \ref{as1}  and the choice of $m$ in the statement,  we can assume that %\red{[L: not very clear the conseuqence from Assumption 1?. btw, is it possible that its $\Delta \geq 3cT/2m$, with the c being the same $c$ as below. This would gurantee that in the set before and after there are no change point if in $\Lambda_j$ there is one]} \blue{Oscar: I changed the wording of Assumption 1, basically  for $c_1$ large enough we can guarantee it}
	\[
	\frac{3 T}{m }\,\leq \,  \Delta.
	\]
	
	Also,  by definition of  $\tilde{Y}$, it holds that
	\[
	\tilde{f}_j \,=\,  \frac{1}{  \sqrt{\vert \Lambda_j\vert  }  }\sum_{i  \in \Lambda_j } \mathbb{E}\left(Y_i\right)\,=\,\frac{1}{  \sqrt{\vert \Lambda_j\vert  }  }\sum_{i  \in \Lambda_j } \sum_{l=1}^i  \Delta f_l . \,\,\,
	\]
	Hence, $	\tilde{f}_j = 	\tilde{f}_{j-1}$ if  $\mathcal{C}^*  \cap \Lambda_{j-1} =\emptyset$  and $\mathcal{C}^*  \cap \Lambda_j =\emptyset$.

	Next, write  $\Lambda_{j} =   [a,  a+T/m ] \cap \mathbb{N}$ and  $a + \eta   \in \Lambda_{j}\cap \mathcal{C}^*$. Also, suppose that  $\eta \geq  T/(2m)$. 
	Then
	\[
	\begin{array}{lll}
		\vert  	\tilde{f}_{j+1} -	\tilde{f}_{j}\vert &=&\displaystyle \left\vert   \frac{1}{  \sqrt{\vert \Lambda_{j+1}\vert  }  }\sum_{i  \in \Lambda_{j+1}  } \sum_{l   =1}^i \Delta f_l    - \frac{1}{  \sqrt{\vert \Lambda_{j}\vert  }  }\sum_{i  \in \Lambda_{j}  } \sum_{l   =1}^i   \Delta f_l  \right \vert\\
		& = &	\displaystyle   \left\vert     \sqrt{ \frac{T}{m} } \,  \left(   \Delta f_1+   \ldots+ \Delta f_{a+\eta} \right)   - \frac{  \eta  }{  \sqrt{\frac{T}{m}  }  } \left(   \Delta f_1+   \ldots+ \Delta f_{a+\eta-1} \right)      -  \frac{  \frac{T}{m} - \eta  }{  \sqrt{\frac{T}{m}  }  } \left(   \Delta f_1+   \ldots+ \Delta f_{a+\eta} \right)   \right \vert\\
		&=  & \displaystyle \frac{\eta  \,\vert \Delta f_{a+\eta}\vert  \,\sqrt{m} }{  \sqrt{ T} }\\
		&\geq& \displaystyle\sqrt{  \frac{T}{m} }\,\frac{\kappa}{2}.
	\end{array}
	\]
	If    $\eta \leq  T/(2m)$, we have that
		\[
	\begin{array}{lll}
	\vert  	\tilde{f}_{j} -	\tilde{f}_{j-1}\vert &=&	\displaystyle   \left\vert   \frac{  \eta  }{  \sqrt{\frac{T}{m}  }  } \left(   \Delta f_1+   \ldots+ \Delta f_{a+\eta-1} \right)      +  \frac{  \frac{T}{m} - \eta  }{  \sqrt{\frac{T}{m}  }  } \left(   \Delta f_1+   \ldots+ \Delta f_{a+\eta} \right)  - \sqrt{ \frac{T}{m} } \,  \left(   \Delta f_1+   \ldots+ \Delta f_{a+\eta-1} \right) \right \vert\\
		&=  & \displaystyle \left\vert \sqrt{ \frac{T}{m} } \, \Delta f_{a+\eta} -\frac{  \eta  }{  \sqrt{\frac{T}{m}  }  } \Delta f_{a+\eta}  \right\vert\\
	&\geq& \displaystyle\sqrt{  \frac{T}{m} }\,\frac{\kappa}{2}.
	\end{array}
	\]

%	\[
%	\vert  	\tilde{f}_{j} -	\tilde{f}_{j-1}\vert \,\geq \, \displaystyle\sqrt{  \frac{cT}{m} }\,\frac{\kappa}{2}.
%	\]

\end{proof}

\section{Proof of Theorem  \ref{thm1} }

%\gr{[change $\mu$ con $f$ (or $\tilde{f}$) if I got it right, actually, I am thinking that the $\beta$ could be left as $\Delta f$ ? (discuss)]}

Theorem  \ref{thm1}  is a consequence of Theorem 4.1 in \cite{nar14},  who prove that the posterior probability of the true linear regression model goes to one as the sample size goes to infinity ($P(\textbf{Z}=t|\textbf{Y},\sigma^2) \overset{P}{\to} 1$ as the sample size goes to infinity).  To use their result, we  verify that the Bayesian model in \eqref{eqn:bayesian} satisfies the conditions in \cite{nar14}. However, model \eqref{eqn:bayesian} relies on the ``transformed data" $\widetilde{\textbf{Y}}$ to estimate $\widehat{\mathcal{C}}$, whereas the linear regression model in Theorem 4.1 \cite{nar14} employs directly the data \textbf{Y}. This difference is the reason why our statement has a localization rate instead of the posterior of \textbf{Z}. This enters into the proof checking the identifiability of the active coefficients (the change points in our case) of the underlying linear model. Lemma~\ref{lem2} defines how the localization rate is achieved through the data transformation.  

%\red{Theorem 4.1. conditions are tied to the linear regression design. Condition 4.1 is a condition on the number of columns of $X$, the design matrix. Condition 4.2 is a condition on the }

Throughout the proof, $\tilde{X}$ denotes the matrix  $X(m) \in \mathbb{R}^{m \times m}$ as in  (\ref{eqn:matrix}), and we use the notation from Lemma \ref{lem2}.  With such notation we write  $\Delta \tilde{f}_1   = \tilde{f}_1$  and  $\Delta \tilde{f}_j =  \tilde{f}_{j} -\tilde{f}_{j-1}$  for  $j \in \{2,\dots,m\}$. Hence, 
\[
\mathbb{E}(  \tilde{Y})   \,=\,   \tilde{X} \bm{\Delta \tilde{f}}. 
\]
Condition 4.1 in \cite{nar14} defines a bound on the total number of covariates, \textit{i.e.} the columns in the design matrix. We are not in a high-dimensional setting (number of covariates much larger than the sample size), hence, the condition is readily met.  In relation to the notation used in \cite{nar14}, we have $p_m = m-1$. Then,
\[
\frac{\log (p_m)}{m} \,=\,   \frac{\log(m-1)}{m} \,\rightarrow \,0
\]
since  $m \rightarrow \infty$ by Assumption \ref{as1}. %\sout{Hence, Condition 4.1 from {nar14} is met.}

Condition 4.2 in \cite{nar14}  imposes restrictions to the prior hyperparameters. It is satisfied by choosing   $\tau_{0,m}^{2}= o(1/m)$, $q_m  \asymp  1/m$ , and $\tau_{1,m}^2  \asymp      m^{1+3\delta}$ for some  $\delta>0$.

In Condition 4.3 \cite{nar14} assumes a fixed number of active covariates. They allow for inactive covariates having nonzero coefficients as long as these are small enough. Our assumptions are more restrictive given that we also have a fixed number of change points but we do not allow for arbitrarily small jumps in $f_t$ at non change points. Hence, Condition 4.3 holds immediately. %\gr{[because we fix all of them to zero by assumptions? or can we incorporate the fact that they could be slightly bigger but small?]} \textcolor{red}{we treat therm as zero}.    

Next we verify  Condition 4.4 from \cite{nar14}.  This refers to the identifiability of the linear model. To verify such condition we rely on Lemma~\ref{lem2} which characterizes the jump sizes in the transformed data $\widetilde{\textbf{Y}}$.  In words, the data transformation allows us to enhance the signal $\Delta \tilde{f}$ which  leads to an identifiable model at the prescribed localization rate. Condition 4.4 amounts to showing that there exists a $N>0$ such that  $N>  1+ 8/\delta$  such that 
%	 The localization rate implied by Lemma~\ref{lem2} is included in this condition.
	%  It shows how the data transformation $\widetilde{\textbf{Y}}$ affects the localization of the change point through the partition size $m$. In words, the data transformation allows us to enhance the signal $\Delta \tilde{f}$ and leads to an identifiable model at the prescribed localization rate. Condition 4.4 amounts to showing that there exists a $N>0$ such that  $N>  1+ 8/\delta$  such that %$\Delta_m(N ) >   5\sigma^2 \vert  \widetilde{\mathcal{C}}\vert(1+\delta) \log\left( m  \right)$, where $\widetilde{\mathcal{C}}\,= \,\{j\,:\,  \tilde{\beta}_j \neq 0   \}$
%\red{[L:note, removed the commented part. seems like $|\widetilde{\mathcal{C}}|=K$, so (18) was repeated twice]  \blue{Othanks!} }
\begin{equation}
\label{eqn:check}
\Delta_m(N ) >   5\sigma^2 K(1+\delta) \log\left( m  \right), %\gr{[\text{for me, I miss where the 2 factor comes from}]}
\end{equation}
where $K$ is the number of $\Delta \tilde{f}_j>0$, \textit{i.e.} the number of change points,  and
\[
\Delta_m(N)  \,:=\,\underset{k \,: \vert k\vert<  N\, K ,\,\widetilde{\mathcal{C}} \cap k^c \neq \emptyset   \, }{\inf}\,  \|(I- P_k) \tilde{X}_k  \bm{\Delta \tilde{f}} \|^2,
\]
where $\tilde{X}_k$  is  submatrix of $\tilde{X}$  consisting of the columns of  $\tilde{X}$ in $k$, $P_k$ is the projection matrix onto the column space of $\tilde{X}_k$, and $\widetilde{\mathcal{C}}:=\{j: \Delta \tilde{f}_j>0\}$.
%\sout{Hence, it is enough to check that }
%\begin{equation}
%	\label{eqn:check}
%	\Delta_m(N ) >   5\sigma^2 K(1+\delta) \log\left( m  \right). %\gr{[\text{for me, I miss where the 2 factor comes from}]}
%\end{equation}

However, as in Remark  4 from \cite{nar14}, we have that
\begin{equation}
	\label{eqn:lower}
	\begin{array}{lll}
		\Delta_m(N)&\geq & \displaystyle  m\,\|\Delta \tilde{f}\|^2   \lambda_{\min}\left(      \frac{\tilde{X}^{\top}   \tilde{X}}{m}  \right)\\
		&   \geq &   \displaystyle    K \frac{T \kappa^2  }{4}     \lambda_{\min}\left(      \frac{\tilde{X}^{\top}  \tilde{X}}{m}  \right)\\
		&   \geq &   \displaystyle     K  \frac{T \kappa^2  }{16m }, \\
	\end{array}
\end{equation}
where the second inequality follows by Lemma \ref{lem2}  and the third  one by Lemma \ref{lem1}. Therefore, (\ref{eqn:check})  holds provided that
\[
 K \frac{T \kappa^2  }{16m}  \,\geq \,  5\sigma^2  K(1+\delta) \log\left( T  \right)
\]
which holds if 
\[
\frac{\kappa^2 T}{\log T}   \,\geq \,  80  \sigma^2\left(1+\delta\right) m
\]
and this holds  if we take $\delta \in (0,2]$, $N>5$, and with $m $ as
\[
m \,=  \,    \floor*{  \frac{\kappa^2   T}{160 \sigma^2   \log T } }.
\]

We now proceed to  verify that Condition 4.5 in \cite{nar14}  holds. Condition 4.5 defines constraints on the minimum and maximum eigenvalues of the Gram matrix. Here we have a fixed design matrix, whose eigenvalues have been studied in Lemmas \ref{lem1} and \ref{lem3}.  Let $a$ be a constant satisfying  $0<a <  (N-1)/\delta$. Then, from Lemma \ref{lem3}, for the first part of Condition 4.5 from \cite{nar14} to hold  it is enough to have that
\[
2m \,<\, \max\{       (m \tau_{0,m}^2)^{-1} ,   m\tau_{1,m}^2   \}, 
\]
but this holds by our choice of  $\tau_{0,m}$ in the statement of Theorem \ref{thm1}. For the second part of  Condition 4.5, by Lemma \ref{lem1},  it is enough to have that
\[
\lambda_{\min}\left(  \frac{    \tilde{X}^{\top}\tilde{X}  }{m} \right) \,\geq \,\frac{1}{4m} \,\geq \,\max\{  (m-1)^{-a},  (m-1)^{-\delta} \}
\]
which holds if  $\delta>1$ and  $a>1$. %\red{[L: I am still verifiny whether the change in constant as an impact, but the above is necessary? I mean do we need the $a>1$, the two arguments of the maximum are identical except that they used diffference parameters]}
%\blue{yes, otherwise the right hand side could be larger.}

%\red{[L: still to add, a paragraph where we comment a bit more on how the statement of Theorem 1 (e.g haussdorf metric and so on) follows from all Z. Thereom 4.1 is mostly on the Z's. I guess just in words but would be nice]}
\section{Proof of Theorem \ref{thm2}}

%\gr{[$\beta_{j0}$ not defined?]. I could be useful to rewrite the theorem in the new notation? (also valid for previous section) }

\begin{proof}
	First, without loss of generality, let us assume that  $\Delta f_{j_0}>0$.  Next notice that 
	\begin{equation*}
		\label{eqn:first}
		\begin{array}{lll}
			\tilde{H_j} &=&   X_{-j}\left(       \frac{\tau_T^2     X_{-j}^{\top} X_{-j} +I   }{\tau_T^2}      \right)^{-1} X_{-j}^{\top}\\
			&  =& \tau_{T}^{2}X_{-j} \left(    \tau_T^{2}     X_{-j}^{\top} X_{-j}+I      \right)^{-1}X_{-j}^{\top}\\
			& \underset{\tau_T \rightarrow0}{\rightarrow}& 0.
		\end{array}
	\end{equation*}
Hence,  for all  $j $ it holds that  
\begin{equation}
	\label{eqn:first_2}
  \underset{\tau_T  \rightarrow 0  }{\lim}	 \,\frac{     x_j^{\top}(I- \tilde{H}_j)x_{j_0} \Delta f_{j_0}   } { x_j^{\top}(I- \tilde{H}_j)x_j +   \tau_{1,T}^{-2}  }  \,=\,  \frac{    x_j^{\top} x_{j_0}\Delta f_{j_0}      }{   x_j^{\top} x_{j}  +\tau_{1,T }^{-2}}.
\end{equation}

Next, let  $j  \neq j_0$ such that  $   \min\{T-j,j\}   \,\geq \, c T$.  Then from (\ref{eqn:first_2}),   for small enough  $\tau_T$ we have that  for all $j $,
	\begin{equation}
		\label{eqn:second}
		\begin{array}{lll}
			\displaystyle 	   \,\mathbb{E}(\mu_{1,j}) & = &  	   \displaystyle 	\,   \frac{     x_j^{\top}(I- \tilde{H}_j)x_{j_0} \Delta f_{j_0}   } { x_j^{\top}(I- \tilde{H}_j)x_j +   \tau_{1,T}^{-2}  }\\
			& =&     	  \displaystyle 	\,\,\frac{    x_j^{\top} x_{j_0}\Delta f_{j_0}      }{   x_j^{\top} x_{j}  +\tau_{1,T }^{-2}} +   \varepsilon_{T,j}\\
			& =&   \displaystyle 	\,\,\frac{    \min\{T-j,T-j_0\}\Delta f_{j_0}      }{T-j  +\tau_{1,T }^{-2}}+\varepsilon_{T,j}
		\end{array}
	\end{equation}
	where  $\vert \varepsilon_{T,j} \vert \leq \varepsilon_T$  for all $j$, with $\varepsilon_T \rightarrow 0$ independent of  $j$, and that can be chosen small enough based on the choice of $\tau_T$ and by  (\ref{eqn:first_2}). Here, we have also used the fact that  $x_j^{\top} x_{j^{\prime}} =  \min\{T-j,T-j^{\prime}\}$  for all $j,j^{\prime} \in \{1,\ldots,T\}$.
	% Furthermore,  if  $\tau_n$ is small enough, $\epsilon_n$  also satisfies $\vert  \epsilon_{n,j}^{\prime}\vert \leq \epsilon$ for all $j$, where  
	%\begin{equation}
	%\label{eqn:third}
	%\begin{array}{lll}
	%	\displaystyle 	   \,\mathbb{E}(\mu_{1,j_0}) & = &  	   \displaystyle 	\,   \frac{     x_{j_0}^{\top}(I- \tilde{H}_{j_0})x_{j_0} \beta_{j_0}^*   } { x_j^{\top}(I- \tilde{H}_{j_0})x_{j_0} +   \tau_{1,n}^{-2}  }\\
	%& =&     	  \displaystyle \,\,\frac{    x_{j_0}^{\top} x_{j_0}\beta_{j_0}^*      }{   x_{j_0}^{\top} x_{j}  +\tau_{1,n }^{-2}}    \,+\,      \epsilon_{n,j}^{\prime} \\	
	% & =& \displaystyle 	\,\,\frac{   (n-j_0)\beta_{j_0}^*      }{n-j_0  +\tau_{1,n }^{-2}}\,  +\,      \epsilon_{n,j}^{\prime}
	%\end{array}
	%\end{equation}
	Therefore, for  $j\neq j_0$  we have that 
	\[
	\begin{array}{lll}
		\displaystyle 	   2\left[\mathbb{E}(\mu_{1,j_0})  - \mathbb{E}(\mu_{1,j})\right] &=&	\displaystyle 	2\left[\frac{   (T-j_0)\Delta f_{j_0}      }{T-j_0  +\tau_{1,T }^{-2}}- \frac{    \min\{T-j,T-j_0\}\Delta f_{j_0}      }{T-j  +\tau_{1,T }^{-2}}\right] +   2\varepsilon_{T,j }-2 \varepsilon_{T,j_0 }\\
		&\geq & \displaystyle 	 2\left[\frac{   (T-j_0)\Delta f_{j_0}     }{T-j_0  +\tau_{1,T }^{-2}}- \frac{    \min\{T-j,T-j_0\}\Delta f_{j_0}      }{T-j  +\tau_{1,T }^{-2}}\right]   -  4 \varepsilon_T  \\
			&\geq & \displaystyle 	 \left[\frac{   (T-j_0)\Delta f_{j_0}     }{T-j_0  +\tau_{1,T }^{-2}}- \frac{    \min\{T-j,T-j_0\}\Delta f_{j_0}      }{T-j  +\tau_{1,T }^{-2}}\right]   \\
		& = :&  \Psi(j),
		% -2\epsilon_{n,j}  
	\end{array}
	\]
	where the second inequality holds provided that
	\[
	4 \varepsilon_T  \,\leq\,  \underset{j \neq j_0}{\min}	 \left[\frac{   (T-j_0)\Delta f_{j_0}     }{T-j_0  +\tau_{1,T }^{-2}}- \frac{    \min\{T-j,T-j_0\}\Delta f_{j_0}      }{T-j  +\tau_{1,T }^{-2}}\right]  
	\]
	which is possilbe by choosing  $\tau_T$ small enough since $\varepsilon_T  \rightarrow 0$ as 
	 $\tau_T \rightarrow 0$. Now notice that if  $j \geq j_0$  then 
	\begin{equation}
		\label{eqn:lower1}
		\begin{array}{lll}
			\Psi(j)   & =  &     \displaystyle \frac{ (j-j_0)\Delta f_{j_0}  \tau_{1,T}^{-2}}{(T-j +\tau_{1,T }^{-2})(T-j_0  +\tau_{1,T }^{-2})}\\
			& \gtrsim& \displaystyle \frac{(  j-j_0)   \Delta f_{j_0} }{  \tau_{1,T }^2  T^2  }.
		\end{array}
	\end{equation}
	Furtheremore, if  $j<j_0$, then
	\begin{equation}
		\label{eqn:lower2}
		\begin{array}{lll}
			\Psi(j)   & =  &     \displaystyle \frac{ (j_0-j)\Delta f_0 (T-j)}{(T-j +\tau_{1,T }^{-2})(T-j_0  +\tau_{1,T }^{-2})}\\
			& \gtrsim& \displaystyle \frac{( j_0-j)   \Delta f_{j_0} }{T}.
		\end{array}
	\end{equation}
	
	Next  denote  $\epsilon \,:=\,  Y- X\Delta f $ and 
	notice that
	\[
	\begin{array}{lll}
		\Delta_j\,:=\, 	(\mu_{1,j_0}-   \mathbb{E}(\mu_{1,j_0})) - 	(\mu_{1,j}-   \mathbb{E}(\mu_{1,j})) & = &    \displaystyle 	\,\,\frac{    x_{j_0}^{\top} \epsilon     }{   x_{j_0}^{\top} x_{j_0}  +\tau_{1,T }^{-2}} - \frac{    x_{j}^{\top} \epsilon     }{   x_j^{\top} x_{j}  +\tau_{1,T }^{-2}}  \,-\, r_{T,j}\\%documentclass[options]{class}\
		% & =& \displaystyle 	\,\,\frac{    x_{j_0}^{\top} \varepsilon     }{   x_{j_0}^{\top} x_{j_0}  +\tau_{1,n }^{-2}} - \frac{    x_{j}^{\top} \varepsilon     }{   x_j^{\top} x_{j}  +\tau_{1,n }^{-2}}  \,-\, r_{n,j}\\
	\end{array}
	\]  
	where  $r_{T,j}$  is a random sequence that converges to zero as fast as desired by letting  $\tau_T$ being small enough. Hence, if  $j>j_0$,
	\[
	\begin{array}{lll}
		\vert \Delta_j\vert  & \leq &    \displaystyle \bigg\vert 	\,\,\frac{    x_{j_0}^{\top} \epsilon     }{   x_{j_0}^{\top} x_{j_0}  +\tau_{1,n }^{-2}} - \frac{    x_{j}^{\top} \epsilon     }{   x_j^{\top} x_{j}  +\tau_{1,n }^{-2}}\bigg\vert  \,+\, \vert r_{T,j}\vert\\
		& \leq &  \bigg\vert    \sum_{i  =j+1 }^T\epsilon_i   \left[  \frac{1}{T-j_0 +  \tau_{1,T}^{-2}}  -  \frac{1}{T-j +  \tau_{1,T}^{-2} } \right]  \bigg\vert  \,+\,   \frac{  \left \vert    \sum_{i=j_0+1}^{j} \epsilon_i\right\vert  }{   T-j_0 +  \tau_{1,T}^{-2}   }   \,+\, \vert r_{T,j}\vert\\
		% & =& \displaystyle 	\,\,\frac{    x_{j_0}^{\top} \varepsilon     }{   x_{j_0}^{\top} x_{j_0}  +\tau_{1,n }^{-2}} - \frac{    x_{j}^{\top} \varepsilon     }{   x_j^{\top} x_{j}  +\tau_{1,n }^{-2}}  \,-\, r_{n,j}\\
		&    \lesssim & \big\vert     \frac{\sum_{i=j+1}^{T}   \epsilon_i }{\sqrt{T-j}}   \big\vert   \,\frac{  \sqrt{T}  \vert j-j_0\vert   }{  (T-j+  \tau_{1,T}^{-2})(T-j_0 +  \tau_{1,T}^{-2}) }  \,+\,   \frac{\sqrt{ \vert j-j_0 \vert }}{T} \big\vert     \frac{\sum_{i=j_0+1}^{j}   \epsilon_i }{\sqrt{\vert j-j_0\vert}}   \big\vert  \,+\, \vert r_{T,j}\vert\\
		& \lesssim&   \big\vert     \frac{\sum_{i=j+1}^{T}   \epsilon_i }{\sqrt{T-j}}   \big\vert   \,\frac{   \vert j-j_0\vert   }{  T^{3/2} }  \,+\,   \frac{\sqrt{ \vert j-j_0 \vert }}{T} \big\vert     \frac{\sum_{i=j_0+1}^{j}   \epsilon_i }{\sqrt{\vert j-j_0\vert}}   \big\vert \,+\, \vert r_{T,j}\vert.\\
	\end{array}  
	\]
%	\gr{[add comment about the notation]}
	Therefore, by the maximal inequality of Gaussian random variables, there exists a constant $C_1>0$, 
	\begin{equation}
		\label{eqn:error_control}
		\underset{  j\,\,:\,\,    j_0\,\leq j\,\leq T-cT  }{\max}\, \frac{\left\vert \Delta_j \right\vert}{\sqrt{  \vert j-j_0\vert  } }     \,\leq\, C_1  \frac{  \sigma \sqrt{   \log T    } }{T},
	\end{equation}
	with probability approaching one. Similarly,

	With a similar argument  we also obtain that 
	\begin{equation}
		\label{eqn:error_control2}
		\underset{  j\,\,:\,\,    cT\,\leq j\,\leq j_0 }{\max} \frac{\left\vert \Delta_j \right\vert}{\sqrt{  \vert j-j_0\vert  } }     \,\leq\, C_1  \frac{  \sigma \sqrt{   \log T    } }{T},
	\end{equation}
	with probability approaching one. 
	
	Furthermore,  with the same argument from above,  recalling that  $\{ \mu_{1,j}^{\prime}\}$  is the version of  $\{\mu_{1,j}\}$  based on the vector $(-Y_T,\ldots,-Y_1)^{\top}$ , it follows that  for
	\[
	 \Delta_j^{\prime} : =  	(\mu_{1,T-j_0+1}^{\prime} -   \mathbb{E}(\mu_{1,T-j_0+1}^{\prime} )) - 	(\mu_{1,T-j+1}^{\prime} -   \mathbb{E}(\mu_{1,T-j+1}^{\prime} )), 
	\]
	it holds that 
		\begin{equation}
		\label{eqn:error_control3}
		\underset{  j\,\,:\,\,    cT\,\leq j\,\leq T-cT }{\max}\,\frac{\vert \Delta_j^{\prime} \vert}{\sqrt{  \vert j-j_0\vert  } }     \,\leq\, C_1  \frac{  \sigma \sqrt{   \log T    } }{T},
	\end{equation}
	with probablity approaching one. Note that in the reverse data $(-Y_T,\ldots,-Y_1)^{\top}$, the point $T-j_0+1$ corresponds to $j_0$ in the original ``forward" data.

	Then from (\ref{eqn:lower1})--(\ref{eqn:error_control3})  with probability approaching one  for any $j_0 \neq j$,  $cT\leq j \leq  T-cT$,
	  \begin{equation}
	  	\label{eqn:fin}
	  		\begin{array}{lll}
	  		%\underset{  cn\leq j \leq  n-cn }{\min}\,\,
	  		%\displaystyle	\frac{\vert\mu_{1,j_0}\vert + \vert\mu_{1,j_0}^{\prime}\vert }{2}\,-\, \frac{\vert\mu_{1,j}\vert + \vert\mu_{1,j}^{\prime} \vert}{2}  	 & =& 	\displaystyle	\frac{\mu_{1,j_0} -\mu_{1,j_0}^{\prime} }{2}\,-\, \frac{\mu_{1,j} -\mu_{1,j}^{\prime} }{2}  \\
	  		\displaystyle	\frac{\mu_{1,j_0} +\mu_{1,T-j_0+1}^{\prime} }{2}\,-\, \frac{\mu_{1,j} +\mu_{1,T-j+1}^{\prime} }{2}  &  =  &        \displaystyle  \frac{ \mathbb{E}(\mu_{1,j_0} \,-\,  \mu_{1,j}        )  }{2} \,+\, \frac{ \mathbb{E}(\mu_{1,T-j_0+1}^{\prime} \,-\,  \mu_{1,T-j+1}^{\prime}        )  }{2} \,+\, \,\frac{ \Delta_j }{2} \, + \, \,\frac{  \Delta_j^{\prime}}{2}\\
	  		%& &\red{+ \, \,\frac{  \Delta_j^{\prime}}{2}}\\
	  		&\gtrsim &  \displaystyle \frac{ \vert j-j_0\vert   \Delta f_{j_0} }{T}  + \frac{\vert  j-j_0\vert  \Delta f_{j_0} }{  \tau_{1,T }^2  T^2  } \,-\, C_1\frac{\sigma  \sqrt{   \vert j-j_0\vert   \log T    } }{T}  \\
	  		& \gtrsim & \displaystyle \frac{ \vert j-j_0\vert   \Delta f_{j_0} }{T}\\
	  		& >&0,
	  	\end{array}
	  \end{equation}
% \gr{[it gets absorded in the constant, but isn't there are a $2$ in the denominator of the first factor to the right? ]}
	where the 	last  inequality holds  provided that  $\vert j -j_0 \vert \,\geq \,  C_2  \sigma^2\log T/  \Delta f_{j_0}  $  for some large enough  constant $C_2>0$,  giving the desired localization rate.

Finally, we verify that for all $j$ with  $cT\leq  j\leq  T- cT$, it holds that
\begin{equation}
	\label{eqn:final0}
	 \left\vert  \frac{\mu_{1,j} +\mu_{1,T-j+1}^{\prime} }{2} \right\vert  \,=\,  \frac{\mu_{1,j} +\mu_{1,T-j+1}^{\prime} }{2},
\end{equation}
with high probability.  To see this  let
\[
\Delta_j^{\prime\prime}  :=   \frac{\mu_{1,j} +\mu_{1,T-j+1}^{\prime} }{2}  -  \mathbb{E}\left( \frac{\mu_{1,j} +\mu_{1,T-j+1}^{\prime} }{2}\right).
\]
Then by choosing  $\tau_T$ small enough, and defining   $\tilde{\epsilon } = (\epsilon_T,\ldots,\epsilon_1)^{\top}$  with probability approaching one, we have that 
\begin{equation}
	\label{eqn:final}
	\begin{array}{lll}
		\vert \Delta_j^{\prime\prime} \vert  &\leq &\displaystyle \frac{1}{2}  \bigg\vert 	\,\,\frac{    x_{j}^{\top} \epsilon     }{   x_{j}^{\top} x_{j}  +\tau_{1,n }^{-2}} \bigg\vert  + \frac{1}{2}\bigg\vert  \frac{    x_{T-j}^{\top} \tilde{\epsilon}     }{   x_{T-j}^{\top} x_{T-j}  +\tau_{1,n }^{-2}}\bigg\vert  \,+\, \vert r_{T,j}\vert\\
		 & = &\displaystyle \frac{1}{2 }   \bigg\vert   \frac{1}{\sqrt{T-j} }\sum_{l=j}^{T } \epsilon_l \bigg\vert    \frac{  \sqrt{T}  }{T-j+\tau_{1,n }^{-2}}        +   \frac{1}{2  }   \bigg\vert  \frac{1}{\sqrt{j}} \sum_{l=T-j}^{T } \tilde{\epsilon}_l \bigg\vert    \frac{  \sqrt{T}  }{j+\tau_{1,n }^{-2}} \,+\, \vert r_{T,j}\vert\\
		  & \lesssim& \displaystyle  \sigma \sqrt{\log  T} \frac{  \sqrt{T}  }{T-j+\tau_{1,n }^{-2}}   \,+\,\sigma\sqrt{\log  T} \frac{  \sqrt{T}  }{j+\tau_{1,n }^{-2}} \,+\, \vert r_{T,j}\vert\\
		    & \lesssim &\sigma\sqrt{\frac{\log T}{T}}.
	\end{array}
\end{equation}
However,  from (\ref{eqn:second}) it follows that 
\begin{equation}
	\label{eqn:final2}
	\begin{array}{lll}
	\underset{  cT \leq  j\leq  T-cT }{\min}	\,\mathbb{E}\left( \frac{\mu_{1,j} +\mu_{1,T-j+1}^{\prime} }{2}\right)     \gtrsim     \Delta f_{j_0}.
	\end{array}
\end{equation}
Therefore,  (\ref{eqn:final0})  follows combining  (\ref{eqn:final})  and  (\ref{eqn:final2}) and using Assumption \ref{as1}. The conclusion of the theorem follows combining  (\ref{eqn:fin}) with (\ref{eqn:final0}).

%with a similar argument and the fact that $\Delta f_{j_0}>0$  it follows that 
%\[
%\frac{\mu_{1,j} +\mu_{1,T-j+1}^{\prime} }{2}    >0 
%\]
%for all $j$  with probability approaching one, and we arrive at the desired result.
%	The claim follows.
	% and  by our assumption on $\beta_{j_0}^*$. 
	
%	\gr{[Are we missing the argument on the marginal or is it that straightforward? ]}
	
\end{proof}

%\end{document}

\section{Details of the simulation scenarios and the implementations}
\label{app:sim.det}

Below we provide the details of the implementations of each method used in Section \ref{sec:sim}. All results in Section \ref{sec:sim} can be reproduced using the code available at  \texttt{https://github.com/lorenzocapp/solocp\_experiments}. We consider as the change point location the first time instance of a new piecewise constant segment.
\begin{itemize}
	\item \textit{basad.cp}: there is no \texttt{R} package, we used a code kindly made available by \cite{nar14}. The code was developed for a variable selection method. Hence we use a $n\times n$ lower triangular matrix of $1$s as input for the  design matrix. We set $5000$ iterations and a burn-in of $1000$. We tried the method for several $q(0.05,0.1,0.2,0.5)$ and use Algorithm \ref{all:all} to select the change points. 
	\item \textit{ebpiece} \citep{liu20}: there is no \texttt{R} package but the code is publicly available at \url{https://www4.stat.ncsu.edu/~rmartin/Codes/ebpiece.R}. We modified the authors' function \texttt{ebpiece} to include $\widehat{B}$, which are the locations of the change points of the fused LASSO (``one standard error rule") that is used as the initialization. The rest of the parameters are the default parameters suggested by the authors for a similar test signals ($\alpha=0.99,v=2 \widehat{\sigma^2}, \lambda=2$ and $10000$ MCMC iterations). Results are fairly sensitive to $\lambda$: $\lambda=2$ led to the best empirical performance. The following code extracts the change points
	
	\texttt{o <- ebpiece\_mod(y, sig2=sig2hat, 0.99, v=$2 \widehat{\sigma^2}$, lambda=2, M=10000,$\widehat{B}$)\\
	cp <- which(diff(apply(o\$B, 2, mean))>0)+2}
\item \textit{pelt} \citep{killick2012optimal}: We used the \texttt{R} package \texttt{changepoint} on CRAN. Default parameters are used and the change points are extracted with the following code 

\texttt{cp <- cpt.mean(y/mad(diff(y)/sqrt(2)), method="PELT")@cpts}

\item \textit{r-fpop} \citep{fearnhead2018changepoint}: We used the \texttt{R} package \texttt{robseg} available for download at \url{https://github.com/guillemr/robust-fpop}. Default parameters are used and the change points are extracted with the following code

\texttt{res.l2 <- Rob\_seg.std(x = y/sqrt($\widehat{\sigma^2}$), loss = "Outlier",  lambda=2*log(length(y)),\\lthreshold=3*sqrt($\widehat{\sigma^2}$))\\
	cp <- res.l2\$t.est[-length(res.l2\$t.est)]+1}

\item \textit{smuce} \citep{fric14}: We used the \texttt{R} package \texttt{stepR} on CRAN. Default parameters are used and the change points are extracted with the following code

\texttt{cp<-which(abs(diff(fitted(smuceR(y, 1:n, family="gauss"))))>0)+1}

\item \textit{solo.cp} : We developed the \texttt{R} package \texttt{solocp} available for download at\\ \texttt{https://github.com/lorenzocapp/solocp}. Parameters choice is described in Section \ref{sec:sim}. A vignette is included explaining how to use the code. 

\item \textit{wbs} \citep{fryzlewicz2014wild}: We used the \texttt{R} package \texttt{wbs} on CRAN. Default parameters are used and the change points are extracted with the following code

\texttt{w <- wbs(y)\\
	w.cpt <- changepoints(w,penalty="bic.penalty")\\
	cp = sort( w.cpt\$cpt.ic\$bic.penalty)+1}

\end{itemize}

\noindent Below we provide specifications of the test signals $f$ and error terms used in Section \ref{sec:sim_nt1}. 
\begin{itemize}
	\item BLOCKS.out: $K=11$, $\mathcal{C}=\{205, 267, 308, 472, 512, 820, 902, 1332, 1557, 1598, 1659\}$, $T=2048$, and $\bm{\mu}=\{0, 14.64, -3.66, 7.32, -7.32, 10.98, -4.39, 3.29, 19.03, 7.68, 15.37, 0\}$. $\epsilon_t \iidsim 0.95 N(0,\sigma=7) + 0.05 N(0,\sigma=28)$ for $t=1,\ldots,T$.
	\item BLOCKS.gauss: same $K$, $\mathcal{C}=$, $T$, and $\bm{\mu}$ as BLOCKS.out. $\epsilon_t \iidsim N(0,\sigma=7)$ for $t=1,\ldots,T$.
	\item BLOCKS.lap: same $K$, $\mathcal{C}=$, $T$, and $\bm{\mu}$ as BLOCKS.out. $\epsilon_t \iidsim Laplace(\mu=0,\sigma=7)$ for $t=1,\ldots,T$, where $\sigma$ is the dispersion parameter of a Laplace distribution.
	\item BLOCKS.studt: same $K$, $\mathcal{C}=$, $T$, and $\bm{\mu}$ as BLOCKS.out. $\epsilon_t \iidsim \text{Student's t} (0,df=4)$ for $t=1,\ldots,T$, where $df$ is the number of degrees of freedom of a Student's t-distribution.
	
		\item TEETH.out: $K=4$, $\mathcal{C}=\{31,61,91, 121\}$, $T=140$, and $\bm{\mu}=\{0, 1, 0, 1, 0\}$. $\epsilon_t \iidsim 0.9 N(0,\sigma=0.25) + 0.1 N(0,\sigma=1)$ for $t=1,\ldots,T$.
	\item TEETH.gauss: same $K$, $\mathcal{C}=$, $T$, and $\bm{\mu}$ as TEETH.out. $\epsilon_t \iidsim N(0,\sigma=0.25)$ for $t=1,\ldots,T$.
	\item TEETH.lap: same $K$, $\mathcal{C}=$, $T$, and $\bm{\mu}$ as TEETH.out. $\epsilon_t \iidsim Laplace(\mu=0,\sigma=0.3)$ for $t=1,\ldots,T$, where $\sigma$ is the dispersion parameter of a Laplace distribution.
	\item TEETH.studt: same $K$, $\mathcal{C}=$, $T$, and $\bm{\mu}$ as TEETH.out. $\epsilon_t \iidsim \text{Student's t} (0,df=3)$ for $t=1,\ldots,T$, where $df$ is the number of degrees of freedom of a Student's t-distribution.
\end{itemize}

\noindent Below we provide specifications of the test signals $f$ and error terms used in Section \ref{sec:sim_ntlarge}. 

\begin{itemize}
	\item BLOCKS2.out: $K=6$, $\mathcal{C}=\{102, 236, 410 ,666, 829\}$, $T=1024$, and $\bm{\mu}=\{0, 14.64,  -7.32, 3.29, 19.03, 0\}$. $\epsilon_t \iidsim 0.9 N(0,\sigma=7) + 0.1 N(0,\sigma=28)$ for $t=1,\ldots,T$.
	\item BLOCKS2.gauss: same $K$, $\mathcal{C}=$, $T$, and $\bm{\mu}$ as BLOCKS2.out. $\epsilon_t \iidsim N(0,\sigma=7)$ for $t=1,\ldots,T$.
	\item BLOCKS2.lap: same $K$, $\mathcal{C}=$, $T$, and $\bm{\mu}$ as BLOCKS2.out. $\epsilon_t \iidsim Laplace(\mu=0,\sigma=9)$ for $t=1,\ldots,T$, where $\sigma$ is the dispersion parameter of a Laplace distribution.
	\item BLOCKS2.studt: same $K$, $\mathcal{C}=$, $T$, and $\bm{\mu}$ as BLOCKS2.out. $\epsilon_t \iidsim 7 \text{Student's t} (0,df=4)$ for $t=1,\ldots,T$, where $df$ is the number of degrees of freedom of a Student's t-distribution.
	
\end{itemize}

\section{Sensitivity of \textit{solo.cp} to the choices of $\Delta$ and $q$}
\label{app:sensi}

We redo the analysis of Section \ref{sec:sim_nt1} to study the sensitivity of the \textit{solo.cp} algorithm to the choice of parameters $q$, which we recall that can be interpreted as a sparsity inducing parameters, and $\Delta$, which can be interpreted as a way to enforce a minimum spacing conditions between change points. %This study is important because $q$ is known to be one of the most important parameters when applying spike and slab priors to variable selection \citep{nar14,che19}, while $\Delta$ is the only parameter not studied in Section THEORY.
Tables \ref{tab:delta} summarizes $|\widehat{\eta}-\eta|/K$, $|\eta-\widehat{\eta}|/\widehat{K}$, $K-\widehat{K}$, $d(\widehat{\mathcal{C}},\mathcal{C})$, and the average computation time for the four scenarios considered for the BLOCKS test signal, varying the parameter $\Delta$ and a fixed $q=0.1$. Table \ref{tab:q} is an identical table where we report the result for a fixed $\Delta=5$ and a varying $q$. The other parameters of the \textit{solo.cp} algorithm are set as in Section \ref{sec:sim_nt1}. 

The robustness of the algorithm to parameter choice is striking, being the sensitivity to these two parameters minimal. The criteria $|\widehat{\eta}-\eta|/K$, $|\eta-\widehat{\eta}|/\widehat{K}$, and  $d(\widehat{\mathcal{C}},\mathcal{C})$ are practically identical within a data type as $\Delta$ and $q$ vary. The bias in the number of change points ( $K-\widehat{K}$) is the quantity more affected by these parameters. As $\Delta$ grows, \textit{solo.cp} moves from overestimating the number change points ($\widehat{K}>K$) to underestimating it (Table~\ref{tab:delta}). This is expected, given that as $\Delta$ grows, longer time intervals will be classified as ``consecutive". As $q$ decreases, $\widehat{K}$ grows and, in this example, the bias increases (Table~\ref{tab:q}). Again, this is largely expected, given higher values of $q$ lead to a higher probability of being classified as a change point.

\begin{table}[!htbp] \centering 
\caption{\small{\textbf{Appendix \ref{app:sensi}:  Haudorff distance, empirical distributions and estimation bias in $K$ of the procedures considered for the BLOCKS test signals.} Average statistics computed over $100$ simulations for the \textit{solo.cp} algorithm for $q=1$ and a varying $\Delta$. In ``Data", .out refers to mixture of Gaussian errors, .gauss to Gaussian errors, .lap to Laplace errors, and .studt to Student's t errors. For $|\widehat{\eta}-\eta|/K$ and $|\eta-\widehat{\eta}|/\widehat{K}$ the higher the number in the zero column the better. Conversely, for $d(\widehat{\mathcal{C}},\mathcal{C}^*)$ the lower the better. For $K-\widehat{K}$, the closest to the zero the better.}}
	\label{tab:delta} 
	\scalebox{0.8}{
		\hspace{-0cm}\begin{tabular}{@{\extracolsep{5pt}} lc|cccc|cccc|c|c|c} 
			\\[-1.8ex]\hline 
			\\[-1.8ex] 
			& &\multicolumn{4}{c}{$|\widehat{\eta}-\eta|/K$} & \multicolumn{4}{c}{$|\eta-\widehat{\eta}|/\widehat{K}$} & & \\
			Data & Method & $0$ & $1$ & $2$ & $\geq3$ &  $0$ & $1$ & $2$ & $\geq3$  & $K-\widehat{K}$  & $d(\widehat{\mathcal{C}},\mathcal{C})$ & comp. time \\ 
			\hline \\[-1.8ex] 
			\multirow{5}{*}{\rotatebox[origin=c]{90}{BLOCKS.out}}& $\Delta=1$ & 0.4 & 0.15 & 0.06 & 0.38 & 0.37 & 0.14 & 0.06 & 0.42 & -1.42  & 118.28 & 115.39 \\ 
			& $\Delta=3$  & 0.4 & 0.15 & 0.06 & 0.39 & 0.47 & 0.18 & 0.08 & 0.27 & 1.52 & 113.71 & 110.8 \\ 
			& $\Delta=5$  & 0.4 & 0.15 & 0.06 & 0.39 & 0.52 & 0.2 & 0.08 & 0.2 & 2.39  & 108.72 & 112.55 \\ 
			& $\Delta=7$  & 0.4 & 0.15 & 0.06 & 0.39 & 0.53 & 0.2 & 0.09 & 0.18 & 2.64  & 107.4 & 110.25 \\ 
			& $\Delta=9$  & 0.4 & 0.15 & 0.06 & 0.39 & 0.54 & 0.21 & 0.09 & 0.16 & 2.86  & 104.04 & 107.22 \\ 
			\hline
			\multirow{5}{*}{\rotatebox[origin=c]{90}{BLOCKS.gauss}} & $\Delta=1$  & 0.51 & 0.19 & 0.07 & 0.24 & 0.4 & 0.14 & 0.06 & 0.41 & -3.55  & 90.27 & 113.64 \\ 
			& $\Delta=3$  & 0.51 & 0.18 & 0.07 & 0.24 & 0.52 & 0.19 & 0.07 & 0.21 & 0.24  & 86.33 & 108.28 \\ 
			& $\Delta=5$ & 0.51 & 0.18 & 0.07 & 0.24 & 0.56 & 0.2 & 0.08 & 0.16 & 0.99  & 81.38 & 113.58 \\ 
			& $\Delta=7$  & 0.5 & 0.18 & 0.07 & 0.24 & 0.58 & 0.21 & 0.08 & 0.13 & 1.36  & 76.69 & 107.88 \\ 
			& $\Delta=9$  & 0.5 & 0.18 & 0.07 & 0.24 & 0.59 & 0.21 & 0.08 & 0.12 & 1.51  & 73.22 & 104.28 \\ 
			\hline
			\multirow{5}{*}{\rotatebox[origin=c]{90}{BLOCKS.lap}}& $\Delta=1$  & 0.34 & 0.14 & 0.06 & 0.46 & 0.32 & 0.13 & 0.05 & 0.49 & -1.23  & 116.84 & 126.22 \\ 
			& $\Delta=3$  & 0.34 & 0.14 & 0.06 & 0.46 & 0.42 & 0.17 & 0.07 & 0.35 & 1.78  & 111.52 & 124.78 \\ 
			& $\Delta=5$  & 0.34 & 0.14 & 0.06 & 0.47 & 0.45 & 0.18 & 0.07 & 0.29 & 2.67  & 107.91 & 124.1 \\ 
			& $\Delta=7$  & 0.34 & 0.14 & 0.06 & 0.47 & 0.47 & 0.19 & 0.08 & 0.26 & 3.03 & 103.41 & 124.07 \\ 
			& $\Delta=9$  & 0.34 & 0.14 & 0.06 & 0.47 & 0.49 & 0.2 & 0.08 & 0.24 & 3.25  & 100.82 & 115.21 \\ 
			\hline
			\multirow{5}{*}{\rotatebox[origin=c]{90}{BLOCKS.studt}} & $\Delta=1$  & 0.36 & 0.15 & 0.06 & 0.43 & 0.33 & 0.14 & 0.06 & 0.47 & -1.25  & 115.85 & 119.84 \\ 
			& $\Delta=3$  & 0.35 & 0.15 & 0.06 & 0.44 & 0.43 & 0.18 & 0.07 & 0.31 & 1.84  & 111.3 & 114.36 \\ 
			& $\Delta=5$  & 0.35 & 0.15 & 0.06 & 0.44 & 0.48 & 0.2 & 0.08 & 0.24 & 2.7 & 107.07 & 112.32 \\ 
			& $\Delta=7$  & 0.35 & 0.15 & 0.06 & 0.44 & 0.5 & 0.21 & 0.08 & 0.21 & 3.06  & 104.32 & 113.5 \\ 
			& $\Delta=9$  & 0.35 & 0.15 & 0.06 & 0.44 & 0.5 & 0.21 & 0.09 & 0.2 & 3.22  & 101.78 & 111.54 \\ 
			\hline \\[-1.8ex] 
	\end{tabular} }
\end{table}

\begin{table}[!htbp] \centering 
\caption{\small{\textbf{Appendix \ref{app:sensi}:  Haudorff distance, empirical distributions and estimation bias in $K$ of the procedures considered for the BLOCKS test signals.} Average statistics computed over $100$ simulations for the \textit{solo.cp} algorithm with $\Delta=5$ and a varying $q$. In ``Data", .out refers to mixture of Gaussian errors, .gauss to Gaussian errors, .lap to Laplace errors, and .studt to Student's t errors. For $|\widehat{\eta}-\eta|/K$ and $|\eta-\widehat{\eta}|/\widehat{K}$ the higher the number in the zero column the better. Conversely, for $d(\widehat{\mathcal{C}},\mathcal{C}^*)$ the lower the better. For $K-\widehat{K}$, the closest to the zero the better.}}
	\label{tab:q} 
	\scalebox{0.8}{
		\begin{tabular}{@{\extracolsep{5pt}} lc|cccc|cccc|c|c|c} 
			\\[-1.8ex]\hline 
			\\[-1.8ex] 
			& &\multicolumn{4}{c}{$|\widehat{\eta}-\eta|/K$} & \multicolumn{4}{c}{$|\eta-\widehat{\eta}|/\widehat{K}$} & & \\
			Data & Method & $0$ & $1$ & $2$ & $\geq3$ &  $0$ & $1$ & $2$ & $\geq3$  & $K-\widehat{K}$  & $d(\widehat{\mathcal{C}},\mathcal{C})$ & comp. time \\ 
			\hline \\[-1.8ex] 
			\multirow{4}{*}{\rotatebox[origin=c]{90}{\small{B.out}}} & $q=0.05$ & 0.39 & 0.15 & 0.06 & 0.39 & 0.52 & 0.2 & 0.09 & 0.19 & 2.73  & 106.56 & 127.99 \\ 
			& $q=0.1$ & 0.4 & 0.15 & 0.06 & 0.39 & 0.52 & 0.2 & 0.08 & 0.2 & 2.39  & 108.72 & 112.55 \\ 
			& $q=0.2$ & 0.41 & 0.15 & 0.07 & 0.37 & 0.49 & 0.19 & 0.08 & 0.24 & 1.63 & 113.88 & 116.81 \\ 
			& $q=0.5$ & 0.42 & 0.17 & 0.07 & 0.34 & 0.47 & 0.19 & 0.08 & 0.26 & 0.89  & 110.32 & 111.23 \\ 
			\hline
			\multirow{4}{*}{\rotatebox[origin=c]{90}{\small{B.gauss}}} & $q=0.05$ & 0.5 & 0.18 & 0.06 & 0.26 & 0.58 & 0.2 & 0.07 & 0.14 & 1.4 & 81.79 & 128.73 \\ 
			& $q=0.1$ & 0.51 & 0.18 & 0.07 & 0.24 & 0.56 & 0.2 & 0.08 & 0.16 & 0.99  & 81.38 & 113.58 \\ 
			& $q=0.2$ & 0.51 & 0.19 & 0.07 & 0.23 & 0.55 & 0.2 & 0.08 & 0.18 & 0.65  & 82.49 & 114.7 \\ 
			& $q=0.5$ & 0.52 & 0.2 & 0.08 & 0.2 & 0.52 & 0.2 & 0.08 & 0.21 & -0.16  & 83.52 & 114.73 \\ 
			\hline
			\multirow{4}{*}{\rotatebox[origin=c]{90}{\small{B.lap}}} & $q=0.05$ & 0.33 & 0.14 & 0.06 & 0.48 & 0.46 & 0.19 & 0.08 & 0.27 & 2.98  & 107.27 & 138.31 \\ 
			& $q=0.1$ & 0.34 & 0.14 & 0.06 & 0.47 & 0.45 & 0.18 & 0.07 & 0.29 & 2.67 & 107.91 & 124.1 \\ 
			& $q=0.2$ & 0.34 & 0.14 & 0.06 & 0.46 & 0.44 & 0.18 & 0.07 & 0.31 & 2.24  & 110.52 & 129.09 \\ 
			& $q=0.5$ & 0.35 & 0.16 & 0.06 & 0.43 & 0.41 & 0.18 & 0.07 & 0.35 & 1.16  & 112.7 & 126.58 \\ 
			\hline
			\multirow{4}{*}{\rotatebox[origin=c]{90}{\small{B.studt}}} & $q=0.05$ & 0.35 & 0.15 & 0.06 & 0.45 & 0.48 & 0.21 & 0.08 & 0.23 & 3.05  & 106.09 & 129.28 \\ 
			& $q=0.1$ & 0.35 & 0.15 & 0.06 & 0.44 & 0.48 & 0.2 & 0.08 & 0.24 & 2.7  & 107.07 & 112.32 \\ 
			& $q=0.2$ & 0.36 & 0.15 & 0.06 & 0.43 & 0.47 & 0.2 & 0.08 & 0.26 & 2.38  & 109.17 & 118.37 \\ 
			& $q=0.5$ & 0.38 & 0.16 & 0.07 & 0.39 & 0.44 & 0.19 & 0.08 & 0.29 & 1.43  & 115.6 & 115.76 \\ 
			\hline \\[-1.8ex] 
	\end{tabular} }
\end{table}

\bibliographystyle{plainnat}
%\begin{spacing}{1}
	\bibliography{biblio}
%\end{spacing}

\end{document}